# A Unified Intermediate Representation and Execution Architecture for Heterogeneous and Distributed Neural Simulation

**Xinjun Li**

Next Brain

## Abstract

Equation-based neural modeling offers flexibility, but executing models across processors and nodes requires explicit control of order, ownership, and numerical behavior. We present brian2-atlas, a rebuilt execution architecture retaining Brian2's modeling interface and bringing heterogeneous and distributed targets under a common model-to-execution contract. Its unified intermediate representation, AtlasIR (Atlas Intermediate Representation), describes typed state, clocks, schedules, effects, events, and random-stream identities, and is checked by an independent Rust validator. Execution-plan analysis constrains physical transformations; target-specific implementations support Rust CPU execution, CUDA and Apple Metal GPUs, browser WebAssembly, and MPI within declared contracts. Ownership-based event processing, compact topology construction, and partitioned storage connect model preparation to distributed execution. Supported MPI pathways additionally preserve mutable synaptic state, delayed events, and boundary checkpoints across simulation segments. In a repeated full-connectome CPU cohort, the best measured Rust configuration reduced simulation-and-recording time by 5.43-fold relative to the best measured Brian2 C++ configuration on the same Linux host. A 298.9-million-synapse model completed on a 16 GB laptop while the compared C++ preparation path exceeded its controlled memory budget. Distributed execution completed 100.5 seconds of model time for 4,129,924 neurons and 24,126,516,728 recurrent synapses on four nodes and 32 ranks. A five-point synthetic weak-scaling study reached 860 million neurons and 860 billion recurrent connections for 100 ms on 30 hosts and 240 physical worker cores, increasing from 86 million neurons on three hosts. These establish tested capacity configurations rather than statistical equivalence or a matched speed advantage over NEST. Published-model workflows extend validation to explicit NMDA dynamics and contextual dendritic plasticity. A matched eight-core NMDA cohort achieves up to 2.10-fold acceleration, while a common-source dendritic-network retest achieves 1.28-fold acceleration after effect-safe endpoint-expression reuse; historical dendritic cohorts retain their separate source identities. Browser studies demonstrate portable reference execution and a separate full-connectome application. A bounded native-training extension supplies explicit forward and reverse execution plans with surrogate spike gradients. The results characterize semantic, numerical, and resource boundaries rather than universal compatibility or performance superiority.

## 1. Introduction

Neural simulation combines mathematical model development with the practical problem of executing many interacting state updates. Researchers change differential equations, synaptic rules, delays, stimulation protocols, and recording requirements as a study develops. Brian2 supports this process through equation-based model descriptions embedded in Python and code generation for execution [1].

Preserving that modeling interface is valuable when the computational requirements of an experiment change. A model may begin with a small exploratory network, grow to a full-connectome workload, and eventually require more memory or processing capacity than a single machine provides.

Execution across these settings involves more than translating arithmetic expressions. A threshold test before synaptic delivery can observe a different state from the same test after delivery. Reordering synaptic additions changes floating-point accumulation. A delayed event can remain pending after one simulation segment and must still arrive in the next. Splitting a network across MPI ranks changes where state resides and how spikes become visible. Assigning random streams from local indices can change the model instance when the partition changes. These properties are part of the computational experiment and need an explicit representation at the boundary between model preparation and execution.

Several established systems separate model specification from efficient execution. GeNN generates simulation code for graphics hardware 2, Brian2GeNN connects Brian models to that system 3, and Brian2CUDA generates CUDA code through Brian's extensible code-generation infrastructure 4. Brian2Wasm provides a path to browser deployment 12. Distributed neural simulation is also well established: NEST has addressed memory and communication at large machine scales 5, and a multi-area cortical model with approximately 4.1 million neurons and 24 billion synapses has previously been simulated on an MPI–GPU cluster 6. The research question here concerns the execution architecture connecting a retained modeling frontend to these different computational settings.

These extensions already preserve much of Brian2's modeling interface, but deliver complementary targets through separately implemented execution paths. Moving a study among CPU, GPU and browser settings can require coordinating target-specific installation, supported features, precision choices, lifecycle behavior and deployment procedures. These coordination costs are distinct from writing the neuronal equations. A shared Python modeling language therefore does not by itself provide a shared execution contract.

The user-facing objective of brian2-atlas is to minimize this coordination burden for admitted models. Common lowering, model identities, capability checks, execution plans and result integration let target selection operate on a shared model representation. Native CPU, CUDA, Metal and MPI are selected through one Device integration; browser execution uses validated portable artifacts derived from the same semantic boundary. Target-specific compilers and resource requirements remain, but users have a common place to inspect eligibility and execution meaning as their computational setting changes. This is an architectural reduction in integration surfaces; the present study does not measure user time or establish a universally minimal workflow.

We investigate an architecture organized around explicit whole-model execution semantics. The frontend lowers supported models into AtlasIR. An independent validator checks the representation, and logical dependencies constrain target-specific plans. The implementation then controls code generation, storage, scheduling, event handling, communication, and results. This arrangement permits different physical strategies while retaining a concrete object against which their behavior can be checked. It also makes model construction part of the execution design: compact topology recipes can reach a native or distributed builder without first expanding all connections in the Python frontend.

Three contributions structure the study. First, AtlasIR defines an independently checked execution contract covering state domains, schedules, events, effects, and model/run identities; execution-plan analysis uses that contract to constrain transformations and explain physical strategy choices. Second, rebuilt CPU, CUDA, Metal, and WebAssembly execution paths extend platform coverage and permit new performance and memory trade-offs, including optional GPU policy calibration with complete-result checks. Third, a distributed implementation combines topology construction, local ownership, and ordered event exchange to execute networks across nodes, with supported plasticity and boundary continuation adding stateful use beyond static capacity demonstrations. Evaluation connects these contributions through semantic examples, repeated performance measurements, strategy-selection and calibration-cost studies, a memory-constrained cortical microcircuit, a four-node multi-area run, browser deployment, and two published-model workflows. The experiments characterize the measured implementation; they do not assume that every valid AtlasIR model is supported by every target.

## 2. Architecture and semantic contract

### 2.1 Retained frontend and replaced execution core

The implementation enters Brian2 through AtlasDevice, registered by `import brian2_atlas` and selected with `set_device("atlas", engine=...)`, which derives directly from the base Device interface. Brian2 continues to provide model objects, equation and unit processing, symbol resolution, and the abstract update statements produced by its numerical state updaters. One-time frontend initialization can also use Brian's NumPy code objects. These are explicit retained dependencies. The simulation time loop is executed by the new engine, with completed arrays and event records returned to Brian2 state and monitor interfaces.

For a supported network, the device collects model objects and executable statements, lowers them into AtlasIR, validates the resulting document, and constructs a target plan. CPU ahead-of-time compilation produces model-specific Rust code and a separately represented instance. GPU targets construct their own programs, buffers, and dispatch sequences. MPI emits a rank-aware Rust executable with a C communication shim. Generic browser execution uses a WebAssembly build of the reference runtime. Figure 1 separates these paths from the retained frontend. Atlas implements its execution paths directly: its CUDA backend does not delegate to Brian2CUDA, Brian2GeNN or GeNN, and these packages are not runtime dependencies of Atlas. CUDA compilation and execution use the NVIDIA toolchain and driver. This independence concerns the downstream simulation backends; the Brian2 frontend dependencies listed above are retained.

Architectural replacement and source-level compatibility have different meanings. The engine replaces downstream simulation infrastructure for the models it accepts. It does not implement every feature that an arbitrary Brian2 Python program can invoke. Capability checks identify unsupported behavior before execution. Appendix A (Table 7) summarizes feature eligibility across execution targets; Supplement S16 gives the restrictions and evidence interpretation. The implementation is available in the brian2-atlas repository at the fixed commit identified under Code and data availability. Each measurement group records its source, inputs, toolchain and hardware. Qualification applies to the tested capability combinations.

Frontend retention includes compatibility changes to index-dependency ordering, function implementation metadata, and typed expression handling. The retained interface is therefore a modeling foundation with documented adapter and frontend patches, rather than a claim to use an entirely unmodified upstream Brian2 checkout.

CPU coverage has expanded beyond the original point-neuron workloads. Reference and AOT paths support a bounded SpatialNeuron implementation using a semi-implicit cable solver on tree morphologies, point-current inputs, and state recording. This path currently requires f64 and the default spatial schedule. Deterministic f64 NeuronGroup equations also support the declared explicit GSL-method interface and adaptive error control; implicit methods, SpatialNeuron/Synapses GSL combinations, and GPU/MPI GSL execution remain outside that contract. Bounded start/end NetworkOperation callbacks execute between native continuation segments. These are backend-specific extensions, not portable eligibility guarantees for every AtlasIR target.

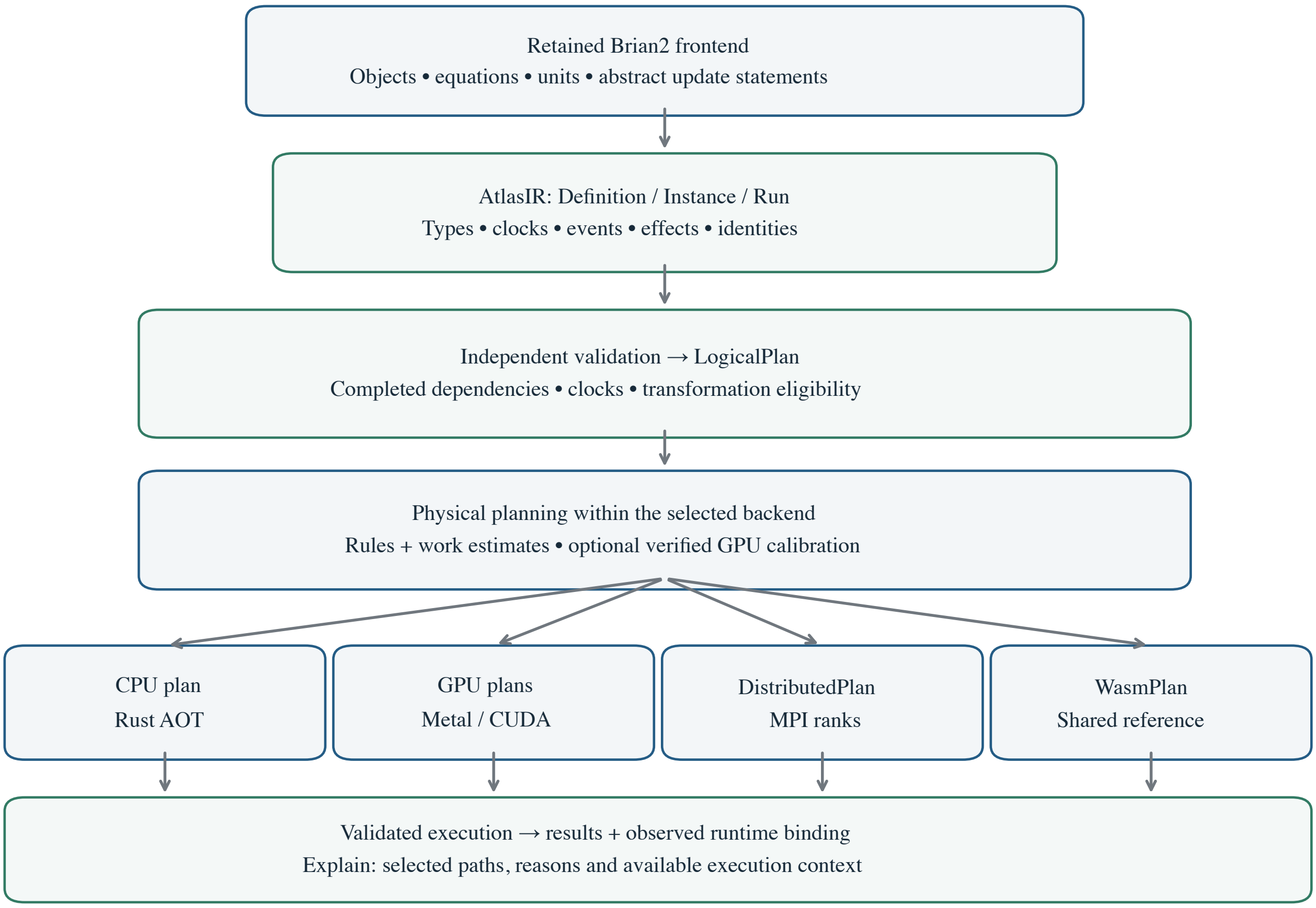

**Figure 1 | Retained modeling, plan analysis and rebuilt execution infrastructure.** Brian2 supplies model preparation and abstract update statements. AtlasIR and independent validation establish the semantic boundary. Completed dependencies constrain physical planning: CPU/GPU planners select supported implementations, while optional GPU calibration compares verified candidates within the selected backend. Plans record decisions and identities; runtime bindings add measured execution context. Generic native reference and WASM share a Rust runtime; CPU AOT, GPU, and MPI use distinct execution implementations. WebGPU and model-specific WASM AOT are separately scoped paths, detailed in Figure 7. The branches denote execution targets integrated in the development tree, while cohort identities and capability checks define what has actually been verified. The architecture does not imply automatic placement across backend families.

## 2.2 AtlasIR: definition, instance, and run

AtlasIR separates a model into three identified layers. The **definition** contains populations, synapses, functions, clocks, executable schedules, and numerical and random-number contracts. The **instance** supplies initial state, parameters, topology or topology recipes, pending events, and the seed. The **run** specifies absolute start time, duration, and each clock's tick interval. Separate hashes distinguish changes in structure from changes in initialized data or requested execution. Reuse of compiled artifacts remains subject to their compatibility rules; independent hashes do not authorize arbitrary instance or run replacement.

Each symbol carries a storage type, a seven-component SI dimension vector, and an index domain such as scalar, neuron, or synapse. Expressions represent loads, casts, arithmetic, comparisons, functions, indexed inputs, and random draws. Update objects include scalar and vector statements, conditional writes, clock membership, execution position, and resource effects. Statement order and old-state snapshots are explicit. The representation contains the discrete update program produced from the frontend equations, rather than requiring the runtime to infer an integration method from an equation string.

Events are named streams. The conventional spike stream, custom events, reset consumers, synaptic pathways, and event monitors have explicit bindings. Pathways carry their own delays and pending events. Refractory behavior is represented through state resources and versioned timing rules. Multiple clocks supply integer tick intervals, with active objects executed in canonical order at the next active time. Edge creation order is retained where it affects event expansion or non-commutative updates.

The wire format uses canonical JSON and SHA-256 layer hashes. Model floating-point values are encoded by their IEEE-754 bits; integer state also uses exact fixed-width encodings. This avoids making model identity depend on decimal formatting or JavaScript number conversion. A consumer verifies the envelope and independently infers expression types and dimensions. Invalid assignments, inconsistent event bindings, forged effects, and malformed schedules are rejected. These checks establish validity under the specification. A hash establishes identity and integrity, not scientific correctness.

### 2.3 Dependencies and observable behavior

The canonical schedule orders executable nodes by slot, order, name, and identifier. Each node has read and write sets. Earlier and later nodes are constrained by read-after-write, write-after-read, and write-after-write conflicts. Consumers also derive implicit effects, notably refractory state accesses associated with thresholds and state updates. Omitting these accesses could make a transformation appear legal even though it changes a subsequent event decision.

A small linked-state example illustrates why dependencies must describe the whole model. One synapse has a clock-driven weight initialized to one and incremented each millisecond. A summed update writes that weight into a target's total variable. A third population reads total through a linked variable and integrates it into x. Under the declared ordering, the start-of-tick monitor records x = [0, 1, 3, 6] over four steps. Deferring the summed update until the end, on the mistaken assumption that only final total matters, leaves the reader's earlier input at zero and produces [0, 0, 0, 0]. The completed read set exposes the intermediate consumer and prevents this transformation (Figure 2).

The example is drawn from a regression comparing reference, AOT, and Brian NumPy execution. It is a deterministic illustration of an optimization constraint, not a statistical benchmark. A planner may fuse or postpone work only when its legality conditions preserve the observable behavior represented by the contract. Otherwise, CPU execution uses a supported canonical slot path or rejects the unsupported configuration. This conservative fallback trades performance for a clear execution meaning.

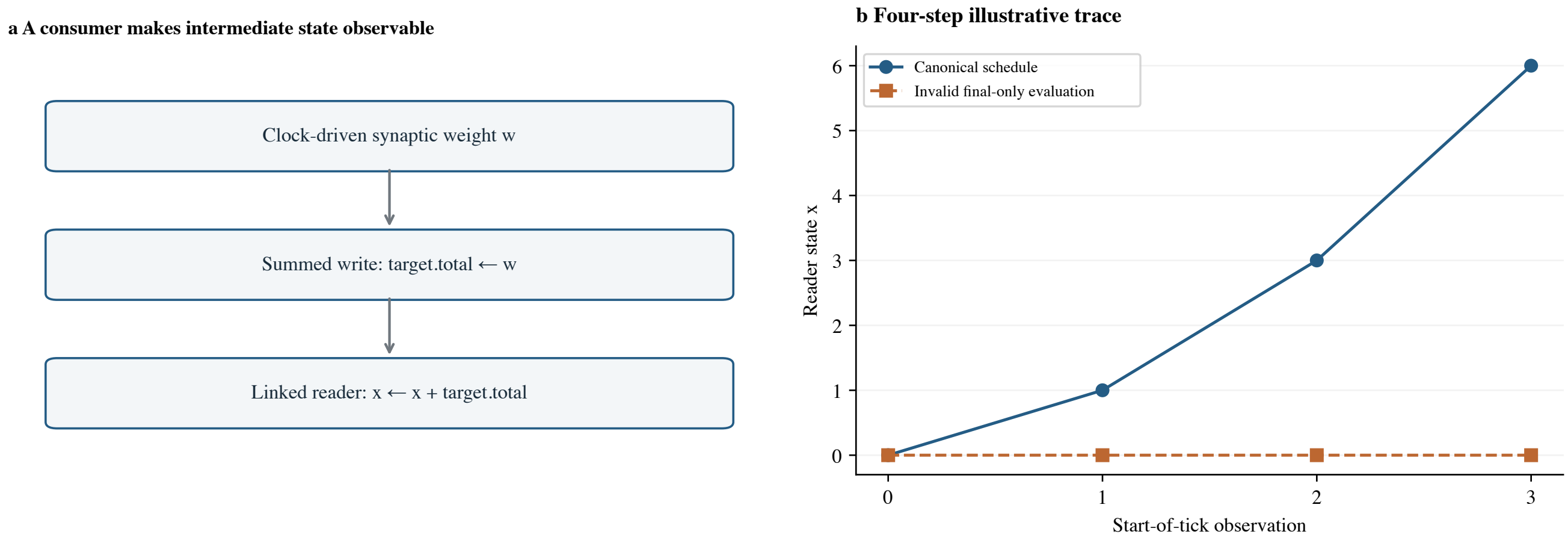

**Figure 2 | Whole-model effects constrain an apparently local optimization.** The summed destination is consumed through a linked read in another population. The sequences illustrate the four-step example and the erroneous final-only transformation. They are explanatory trajectories, not uncertainty estimates.

## 2.4 Execution-plan analysis and semantics-constrained strategy selection

A logical plan contains canonical nodes, completed effects, dependencies, and clock activation. Physical plans select layouts and execution strategies within a target's capabilities. CPU plans choose optimized or canonical generation. GPU plans associate logical work with dispatch stages and buffers. WasmPlan identifies the verified runtime order, and DistributedPlan adds partitions and communication nodes. Before emission or browser execution, the relevant consumer reconstructs and checks the plan instead of trusting arbitrary serialized plan data.

Planning separates whether a transformation is legal from whether a supported implementation is likely to be worthwhile. CPU eligibility checks cover clock activation, schedule contraction, population and synapse effects, linked readers, event pathways, and pending delays. A fixed-phase implementation is accepted only when the checks establish equivalence for the supported schedule. Otherwise, the planner selects the canonical slot implementation described in Section 2.3. Physical choices include compact or general generation, update/threshold fusion, route grouping, target-owned event work, and final-active-tick summed evaluation where intermediate observations permit it. These decisions are consumed by code generation rather than being a descriptive report added after compilation.

Workload heuristics complement the legality checks. The CPU planner estimates generated work from expression operations, assigning greater weight to transcendental functions, division and distribution sampling than to simple arithmetic. It combines this estimate with item counts, indirect synaptic state accesses, minimum useful task sizes and requested parallelism. These proxies determine whether eligible work is split and limit excessive task creation. They are local scheduling heuristics, not a fitted predictor of wall time or a model of total resident memory. GPU planning similarly distinguishes independent temporal fusion from a coupled dependency graph, then uses resource effects to permit edge- or target-owned work while retaining canonical execution for unsupported dependencies.

Canonical CPU summed updates can reuse a pure float64 exponential at the reduction endpoint when it reads only stable endpoint state and scalar inputs. The planner excludes reads of the destination being accumulated, edge-dependent values, vector-rebound inputs, random draws, external function calls and unsupported types, and retains the original edge loop for sparse or small projections. The selected cache is rebuilt after the scalar statements at each original node activation; it does not defer a reduction, reorder edges or change the expression's arithmetic. Its selection, declared payload and node-local lifetime are part of the checked physical plan.

Each plan records the model layer identities, selected policies, declared buffers and dispatch structure. Decisions include a selected path and an explanatory reason, such as insufficient work, an unsafe pathway effect, or a cross-pathway recurrent dependency. The policy identity incorporates the complete logical graph, layout sizes and instance-dependent choices. Consumers re-derive the plan before emission or execution and reject mismatches. Declared array payloads and lifetimes are reported separately from dynamic queues and uninstrumented allocations; the plan inventory does not predict peak RSS or establish general buffer alias reuse.

The public explain_plan interface exposes these decisions before execution, when runtime context is unbound. After a successful run, its runtime binding adds observed thread count, affinity, executed parallel paths and timings when available. Uninstrumented quantities remain explicitly unknown. Optional GPU calibration supplies a second selection mechanism based on complete replays of verified candidates (Section 3.6). Backend selection and numerical mode remain explicit: these mechanisms do not automatically search CPU, GPU, WASM and MPI placements or establish a globally optimal implementation.

The shared representation also supports interactive inspection in the Next Brain workbench. Its AtlasIR inspector presents the ordered schedule, population state and parameters, and a selected node's dependencies, declared reads and writes, and lowered operations. Python-run views inspect the artifact produced by execution. Supplementary Figure S1 pairs saved FlyWire circuit activity with the corresponding catalogue AtlasIR draft under default parameters. The saved activity and model draft have separately stated provenance; the draft is not presented as that run's archived execution artifact. Displayed stages organize logical operations for navigation, rather than exposing the complete backend-specific physical plan or optimization search.

## 2.5 Numerical profiles and execution boundaries

The baseline AtlasIR profile is reference-f64. GPU execution requires an explicit float32 choice, recorded in its target contract even when public arrays are stored as float64. Integer and Boolean state retain their own representations. A wider output container does not restore precision lost during arithmetic. Same-profile conformance, repeatability, and cross-profile comparisons are therefore separate evaluation questions.

**Table 1 | Execution paths and selected boundaries.** This summarizes the described implementation rather than asserting that every combination has been tested. Snapshot details appear in the supplement.

| Path | Implementation and representative behavior | Important boundary |
|---|---|---|
| Native reference | Rust runtime; typed state, canonical schedules, clocks, events, links and monitors | Reference implementation, not an independent scientific oracle |
| CPU AOT | Model-specific Rust; ownership and slot paths; supported plasticity, SpatialNeuron, GSL-method interface and boundary callbacks | Spatial/GSL are bounded CPU f64 paths; callback slots and continuation are constrained |
| Metal / CUDA | Generated GPU programs; independent-cell fusion and coupled networks with declared delays, plasticity, typed state and links | Explicit f32 arithmetic; eligibility and resource limits |
| Generic WASM | Shared reference in a Worker; portable AtlasIR and strict bundle validation | Native-only functions and filesystem CSR rejected; WASM32 limits |
| WebGPU | WGSL after WASM validation; restricted independent-cell f32 models | Experimental; no general synaptic networks, delay queues or STDP |
| Model-specific WASM AOT | Frozen generated Rust with a memory-only host; full-connectome application | Separate specialization, not generic bundle eligibility |
| MPI CPU | Rust AOT/MPI shim; target ownership; explicit/binary topology supports pre/post plasticity, segments and boundary checkpoints | Shared clock; fixed-total procedural remains tick-0 only; no general topology migration or live fault tolerance |
| MPI GPU offload | Explicit CPU/Metal/CUDA rank assignment; CPU f64 and GPU-update f32; supported segmented STDP | Local CPU+Metal tested; host event/synapse/MPI work retained; CUDA simulation offload and cross-host mixed-vendor qualification absent |
| Native training extension | Versioned training plans; CPU/GPU forward and VJP, Rust optimizer, bounded local MPI | Separate plan family; surrogate/discrete gradient contracts; hardware coverage tied to individual snapshots |

## 3. Execution mechanisms

### 3.1 Distributed construction, ownership, and event exchange

MPI assigns each neuron to one rank and each synapse to its target neuron's owner. Local storage contains mutable neuron state, refractory state, applicable parameters, external sources, and incoming connections. Global neuron and edge identities remain available to expressions and random-number generation even though storage indices are local. Thus partitioning changes data placement without redefining model identities.

For static topology, the exporter writes per-rank instance shards and an index describing their sizes and hashes. Generated source refers to metadata instead of embedding the entire network as literals. Each rank checks its communicator and compiled plan, streams its shard into state storage, and validates the bytes actually consumed. Final mutable arrays are gathered from their owners as raw bit patterns rather than combined by floating-point summation, preserving values such as negative zero.

Fixed-total connectivity uses distributed construction. Python emits a compact recipe instead of endpoint and parameter arrays. Ranks process disjoint intervals of global sampling indices. Integer collective operations establish global source-major edge identities, and bounded exchanges route generated edges to target owners. Each rank sorts local connections by the required identity before initialization and execution. Weight, delay, and runtime random draws use the original edge identity. The construction needs temporary storage and sorting; it is not zero-copy generation or constant-memory preparation.

This removes the requirement to materialize all recurrent connections in one Python process. Costs include per-projection source indices, temporary edge records, communication, and local sorting. The documented builder has an O(E_local log E_local) sorting stage and O(N_source) source-index storage per projection. These terms matter for many projections and uneven partitions. Admission limits precede large allocations, but edge counts alone cannot predict total memory because state, queues, recording, and temporary copies also contribute.

During execution, the baseline capacity path exchanges global spike indices after the relevant threshold or event-source node. Collectives occur in canonical order; received indices are sorted and checked. Target ranks expand events against local connectivity. Uniform delays can queue sources and expand them when due; heterogeneous delays retain arrival and enqueue ordering. Message arrival order never substitutes for model event order. Random draws use the model seed, stream, absolute tick, and original neuron or edge identity, without a rank-dependent seed.

The rank-local experiment in Section 5.2 uses all-rank spike exchange at population/tick granularity. It retains replicated read-only presynaptic state, source-offset information, and some monitor layout. These costs help explain why lower local storage does not guarantee strong scaling. Later multi-area implementations use additional construction and recording specializations, with timings bound to their own snapshots. The experiments do not establish general lookahead execution, selective subscription routing, dynamic repartitioning, or recovery from failed cluster nodes.

### 3.2 CPU execution and preparation

CPU AOT compilation specializes state updates and event handlers to the model. State is arranged by variable, avoiding a uniform per-neuron object layout. Eligible worker paths assign writes by target ownership, allowing incoming events to be processed without uncontrolled concurrent floating-point updates to the same target. Work distribution can account for incoming degree, and event routing avoids repeating safely shareable transformations. Small workloads retain serial paths because synchronization can exceed useful work.

Optimized layouts are conditional on dependencies. A canonical slot generator handles supported schedules for which fixed-phase execution cannot establish equivalence. This matters for linked reads, summed variables, custom ordering, and multiple clocks. Adding a consumer that makes an intermediate state observable can therefore change the physical code selected for a model.

Preparation is also specialized. Explicit arrays suit small or already materialized models, while procedural descriptions carry fixed-total topology and initialization into native code. Filesystem-backed CSR represents imported graphs with an explicit content identity. These paths reduce Python materialization and generated-project size in different ways, without a universal low-memory guarantee. The PD14 experiment isolates the practical importance of this preparation boundary on a 16 GB machine.

### 3.3 Metal and CUDA execution

GPU plans distinguish independent cells from coupled networks. For eligible independent populations, a lane can advance one neuron through multiple ticks, reducing dispatch overhead. Coupled models use ordered stages for updates, thresholds, events, synaptic processing, resets, and observation. Stages run independently per edge or target when effects permit; shared writes or other dependencies retain canonical ordering.

Scanning delivery examines ordered incoming connections. Sparse delivery expands fired sources into target queues with integer reservations, then restores canonical order before floating-point updates. Delayed paths account for emission history and arrival order. Integer queue reservations do not imply unordered floating-point accumulation. Suitable policies depend on activity, connectivity, delays, and launch overhead, so the implementation records explicit policy identities.

Metal uses runtime compilation and command-buffer submission on Apple hardware. CUDA has its own compilation, buffer, and launch lifecycle, including retained resources for repeated runs. Reusing compiled programs and immutable topology reduces repeated preparation, while writable state must be reset or restored. Residency, copies, result extraction, and process/file costs all affect the interface. The GPU replay measurements consequently include reset through completed host arrays, rather than reporting only kernel time.

### 3.4 Browser execution and portable artifacts

Generic WebAssembly shares model validation and execution code with native reference. Bundles contain the original model and plan encodings. The browser validates identities and semantics, reconstructs dependencies, and checks dispatch order. Keeping the original encoding avoids losing large-integer precision through JavaScript. After build and delivery, the runtime needs static assets and a browser rather than a Python simulation service.

A Worker isolates execution from the interface and supports batches, progress, and cancellation. Batch size changes host scheduling, not model ticks. Imported bundles and bounded equation authoring are distinct entry points: authoring creates new validated identities, while loading requires supplied identities to match. Portable functions can be represented in AtlasIR; native-only functions and filesystem-backed graphs are rejected by the generic consumer.

Two further paths have separate scopes. Experimental WebGPU generates WGSL for supported independent-cell f32 models. Model-specific WASM AOT compiles a frozen generated Rust simulation with a memory-only host, preserving the application's equations, ordering, random draws, and serialization. The latter runs the full-connectome application in Section 5.5, but does not imply that the generic bundle loader accepts arbitrary full-connectome CSR input. Figure 7 distinguishes these paths.

### 3.5 Recording, continuation, and heterogeneous ranks

Simulation state includes refractory deadlines, queued events, random counters, monitor history, and absolute time as well as neuron arrays. CPU and single-machine GPU lifecycle operations preserve the supported components across runs and checkpoints. Rolling recording retains a bounded observation window while maintaining state needed for continuation; its total memory benefit depends on the surrounding frontend and result representation.

The distributed large-network study uses a separate recording path with bounded spooling and post-run collection. Its completion does not establish checkpoint recovery at that scale. A later explicit/binary-topology path supports segmented plasticity and successful-boundary recovery, as described in Section 3.7; its small-network evidence is separate from the large procedural construction cohort. Optional mixed-device MPI accepts an explicit ordered backend assignment for every rank: cpu, metal, or cuda with a local device ordinal. This supports heterogeneous CPU/GPU rank configurations through a common DistributedPlan, with Metal and CUDA adapters for neuron-state updates. CPU ranks use f64; GPU updates require the declared mixed-f32 profile, with f32 inputs/operations and f64 host-state writeback. Thresholds, resets, synapses, queues, recording and MPI exchange remain on each rank's CPU. Device assignment is part of plan and checkpoint identity, and changes can alter rounding and later spikes. The local Metal experiment and the separately scoped CUDA MPI training record are summarized in Section 5.9. The simulation path requires compatible host OS/CPU ABI and MPI environments; a selector accepting both adapter names does not establish a working macOS-Metal/Linux-CUDA communicator.

### 3.6 GPU policy calibration and verified decision reuse

Optional per-activation autotuning compares four policies within the selected Metal or CUDA backend: baseline, parallel synapse prefix, ordered target bitmaps, and their combination. It retains the requested event-delivery and dependency-graph execution modes. The facility is disabled by default and requires explicit float32 execution. It is a bounded calibration procedure, with no transfer of measured decisions to different models or activity patterns.

Every candidate starts from the same complete initial model, including pending events, absolute clocks and random state. Each candidate is independently planned and validated. Byte-equivalent complete physical plans are deduplicated before another executor is constructed. Distinct candidates can reuse source-identical validated kernels within the activation while retaining independent writable state. Each distinct candidate receives a full warmup and three complete reset-to-result replays, with randomized round order. A fingerprint covers every returned population and synapse array, its dtype and shape, scalar event counts, and numerical and RNG profiles. Every replay must match the first baseline fingerprint exactly. Candidate errors or mismatches exclude that candidate with a recorded reason; baseline failure aborts calibration. This within-backend gate does not establish equivalence to reference-f64.

A candidate replaces baseline only if its median time is at least 5% lower and its slowest measured replay is faster than baseline's fastest. The lowest eligible median wins; otherwise baseline remains selected. The separated-range rule is a conservative short-sample heuristic, not a confidence interval or proof of optimality. The selected executor then performs one additional complete replay, which must pass the same result check. Only this final result updates Brian2 arrays, monitors, clocks and delayed-event state. Thus profiling does not advance the user's simulation repeatedly. Failures in that final run abort publication of the result.

With four distinct candidates, calibration requires up to 17 full executions: four warmups, twelve measurements and one final replay. Preparation, compilation, fingerprinting and cleanup contribute to total calibration cost but are outside the individual replay samples. Up to four executors may coexist, so tuning can require more memory than an ordinary run. A faster selected replay consequently does not imply a faster complete activation.

An optional decision cache stores at most eight metadata records per Device. Keys include the complete input state, topology, pending events, clocks, duration, RNG state, implementation identity and execution options. A tentative hit must also match the independently derived plan and current compiler/device context. A fresh validated executor performs a complete replay and checks its result against the original calibration fingerprint. A mismatch invalidates the record and invokes current-input calibration; interruptions and cleanup failures abort. Only successful frontend result loading publishes a decision. Exact store/restore replay requires restoration of RNG state, whereas changed-state continuation normally misses. The cache holds neither simulation results nor native buffers; compilation and allocation reuse are separate options. It verifies correctness for matching inputs without re-establishing that the cached policy is still fastest.

### 3.7 Distributed plasticity and successful-boundary recovery

For supported explicit and binary-CSR connections, target owners maintain typed mutable synaptic state and execute canonical pre/post pathways. Pair and Triplet STDP include clock-driven or event-driven traces and their last-update state. Post adjacency is ordered by original edge creation identity; delayed pre work retains the declared source and edge order. Writes remain local to the synapse owner and its target neurons. This extension preserves the ownership contract while expanding what state can evolve, rather than relaxing ordering to permit arbitrary remote writes.

Segment continuation retains absolute clocks, refractory state, random-stream identity, monitor history, and pending events. At a successful boundary, owner state and pending work are collected at the coordinator after ranks terminate normally. The disk snapshot is then committed through a checked temporary-file and replacement protocol. Restore validates topology and definition identities, rank and device layout, precision, source/runtime identities, and compiler/platform compatibility before publishing restored arrays. Exact replay requires explicit RNG-state restoration. A failed activation does not commit a new boundary.

Fixed-candidate structural plasticity can change live masks, weights, traces, and birth/activation metadata at a boundary. Generation checks prevent an old queued event from modifying a newly activated edge. Candidate counts and endpoints remain fixed: this is not general CSR growth, repartitioning, or migration. Fixed-total procedural construction retains its single tick-zero activation boundary and rejects mutable-state or pending-event continuation. The checkpoint coordinator and Python host also retain full-state overhead, so recovery is not an O(local edges) distributed storage scheme. These mechanisms provide tested restart behavior for their admitted configurations, not online recovery from failed ranks or cluster nodes.

### 3.8 Bounded native forward and reverse execution

A separate native-training executor accepts versioned plans for dense LIF, explicit recurrent/shared-parameter graphs, scalar equations, multistate equations, and ordered dynamic synaptic transforms. Brian lowering retains units, resolved equations, update/reset order, and canonical parameter/state identities for the accepted subset. CPU or selected GPU kernels execute the forward program and vector–Jacobian products (VJPs); native Rust implements SGD or Adam. Python prepares and transports plans and results. This plan family is an extension of the execution architecture, not a general automatic-differentiation implementation for every valid simulation AtlasIR document.

Hard spikes use a declared surrogate VJP, while Boolean/integer controls and selected discrete gates stop gradients. Coupled updates retain old-state snapshots and reverse the declared update/reset sequence. Dynamic plans include bounded pre/post transforms, delayed event state, summed inputs, and clock-visit ordering. For admitted stochastic programs, pathwise derivatives condition on the sampled trajectory, and Poisson paths use explicit likelihood and boundary-gradient contracts. These definitions do not claim the ordinary derivative of hard spike decisions or unrestricted stochastic autodifferentiation.

Requests can carry forward state and checkpoint parameters, optimizer state, clocks, and supported random/event state. Full BPTT and truncated adjoint windows are distinct contracts; truncation does not by itself reduce the allocated tape. Admission checks bound tape and transport sizes before execution. Training MPI partitions forward/reverse work by target ownership and assigns shared parameters a unique optimizer owner, but currently replicates complete model, input, and tape arrays at each rank. Its local numerical validation consequently establishes neither large-model distributed-memory savings nor cross-host or multi-physical-GPU scaling.

## 4. Evaluation methodology

### 4.1 Workloads and cohorts

The implementation is available in brian2-atlas at commit `ae649a244e28d7d4b3eb5df4e35215e14c1378c7`. Workload drivers, frozen source manifests, inputs and measurement records are archived in brian2-atlas-preprint at tag `biorxiv-v1`. The updated dendritic network cohort re-exports its model and builds from the common-source snapshot documented in that archive. The supplement identifies the exact source version, inputs, toolchain and validation scope for each CPU, GPU, MPI, browser and training cohort.

Evidence is organized into identified cohorts. The CPU full-connectome study uses FlyWire v783, with 139,255 neurons and 15,091,983 directed weighted edges 10. These summarize 54,492,922 contacts, which are not separately simulated edge objects. CPU dynamics are uniform excitatory LIF. The MPI study uses an EI variant with 580 additional input sources and stimulation/cut conditions. Shared graph scale does not make these the same dynamical workload.

The memory study uses a DC-input adaptation of the Potjans–Diesmann microcircuit 9. The large MPI study adapts the multi-area cortical model of Schmidt and colleagues 11. A separate assembly workload is a documented triplet-plasticity variant associated with Litwin-Kumar and Doiron 13, explicitly not a reproduction of all rules in that paper. GPU fixtures isolate delayed STDP and recurrent current-based dynamics. Browser application checks use fixed inputs and a frozen full-connectome executable.

**Table 2 | Principal cohorts and measurement scope.** Durations are model time; repeats belong to the stated comparison rather than a common protocol. Full source details appear in the supplement.

| Cohort | Neurons / simulated edges | Duration | Design | Timed quantity |
|---|---|---|---|---|
| CPU FlyWire | 139,255 / 15,091,983 | 1 s | 1/4/8/16 threads, five repeats after warmup on two hosts | Simulation and recording |
| PD14 laptop | 77,169 / 298,880,968 | Rust 10 s; C++ 0.1 s attempt | Single resource observations | Native body and Device end-to-end separately |
| GPU ring STDP | 4,096 or 16,384 / degree 8 | 256 ticks; 250 ms | Warmup and five randomized rounds | Reset through host results |
| GPU recurrent CUBA | 4,096 / 131,072 | 2,048 ticks | Warmup and five interleaved rounds | Reset through host results |
| GPU policy selection | 4,096 / degree 8 or 32 | 1,024 ticks | Four candidates; one warmup and three randomized replays each; five later winner replays | Profiling time and total calibration separately |
| GPU decision reuse | 4,096 / degree 8 or 32 | 1,024 ticks | Separate cohort; one calibration and three exact-input hits | Activation including transport; excludes frontend lowering |

| Cohort | Neurons / simulated edges | Duration | Design | Timed quantity |
|---|---|---|---|---|
| MPI FlyWire EI | 139,255 + 580 sources / 15,091,983 | 1 s | Before/after, 1/2/4 ranks, three rest repeats and condition checks | Simulation, recording, final gather |
| MPI multi-area | 4,129,924 / 24,126,516,728 recurrent | 100.5 s | Original seed-1729 capacity record; three Rust realizations and three exploratory NEST references | Original launch/resources; later descriptive analysis, no matched speed ratio |
| Connected synthetic MPI weak scaling | 86–860 million / 86–860 billion | 100 ms | 3/6/12/24/30 hosts, eight physical worker cores per host; one observation per size | Preparation and launch through complete output, reported separately |
| Assembly lifecycle | 5,000 / approximately 5 million | Up to 1,000 s recording/recovery | Separate timing, memory and recovery cohorts | Compiled simulation and recovery separately |
| WASM AOT | 139,255 / 15,091,983 | Fixed application protocol | 13 fresh inputs plus browser/offline checks | Conformance; no formal speed claim |
| Explicit NMDA CPU | 2,560–20,480 / 11.8–755.0 million total | 1 s; 10,000 RK4 steps | Matched eight-core Linux; warmup and five/six repetitions | Native initialization, simulation, original recording and result dump |
| Dendritic CPU models | Single-neuron variants and complete Figure-3 topology | 10 s or admitted 100 ms | Cython/Rust/Rust/Cython order; ten or six measured samples per backend | Steady-state simulation and required recording |
| MPI plasticity / native training | Bounded fixtures; sizes vary by contract | Segmented and optimizer-step protocols | Snapshot-specific numerical and restart tests | Functionality; no training throughput or accuracy benchmark |

## 4.2 Correctness and numerical comparison

Structural checks verify IR, identities, types, dimensions, schedules, and plans. Execution checks compare final state, trajectories, event identifiers and times, and lifecycle state. Application checks concern the interpretation of model outputs. Structural validity does not establish biological correctness, and matching implementations can leave shared specification or frontend errors undetected.

Where the numerical program permits it, comparisons use exact array or file equality. CPU FlyWire checks all final voltages, refractory state, spike identifiers/times/counts, and voltage trajectories from 16 neurons. MPI rank-local checks complete result and event files against a same-source native reference. GPU studies use explicit f32 controls and fixture-specific tolerances, retaining f64 diagnostics separately. Recurrent CUBA requires exact spike ticks/indices and final-voltage tolerances rtol 1e-4 and atol 5e-6 for its matched f32 gate. Failed qualification is reported and excluded from the corresponding timing ranking.

Replay tests assess repeatability from an initial instance. Continuation tests assess segment boundaries. Fresh-process recovery tests assess loading saved state in a new process. Browser batch invariance checks the same WASM program under different host batches. These are distinct questions; none establishes universal cross-platform bitwise equivalence. Application comparisons with different stochastic streams require statistical interpretation rather than per-event identity.

### 4.3 Timing and memory

CPU FlyWire measures simulation and recording, excluding preprocessing, compilation, executable startup, and final file writing. Four thread counts are measured with five interleaved repeats after warmup. Ratios compare medians within a host. Both matched-thread results and each backend's best measured configuration are retained. The bounded configuration search does not establish a global optimum.

GPU workers compile once, produce bootstrap output, warm up, and execute five randomized rounds sequentially per device. The timer starts at reset/initialization and ends with complete host results. Depending on the adapter, this includes standalone process launch, output files, readback, copies and load/unload. Compilation, export, hashing and archive serialization are excluded. Large replay ratios can therefore reflect lifecycle differences. The serial Rust slot executor and compiled f32 expression control in the ring cohort are not tuned multicore CPU baselines.

Native-child RSS excludes frontend and compiler memory. Maximum per-rank RSS is not total cluster memory. Summed process-tree peaks are a different metric, and independent node peaks cannot be summed into a simultaneous cluster peak. GPU buffer limits and WASM linear memory exclude other allocations. Resource tests distinguish controlled termination during severe swapping from an observed operating-system OOM kill.

### 4.4 Baselines and provenance

The original CPU comparisons use Brian2 C++ standalone; the additional dendritic cohort uses Brian2 Cython runtime and is reported separately. GPU comparisons include Brian2CUDA, Brian2GeNN, and direct GeNN, with stock and modified adapters distinguished. These systems are used as separately configured benchmark comparators, not as execution components of Atlas. Corrected scheduling or arithmetic variants are not presented as stock packages. NEST provides distributed comparison context, but this version does not report a completed matched multi-area speed ratio. Historical runs with differing outputs, resources, or tuning are not divided to manufacture one.

The supplement maps cohorts to reports and source identities. For several CPU and lifecycle cohorts, verification covers retained reports; full raw arrays remain outside the accompanying evidence package. MPI rank-local and GPU population-scale figure data were additionally extracted from machine-readable reports. Figures distinguish report-derived summaries from individual recorded samples. Missing samples and uncertainty intervals are not reconstructed.

### 4.5 Policy-selection and calibration-cost protocol

The selection study uses two delayed-STDP workloads on M3, L4 and A100: 4,096 neurons, 1,024 ticks with dt = 1/1,024 s, and fixed outdegree 8 (quiet) or 32 (wide). Topology uses seed 42. Quiet uses drive 1/64, pre delays distributed over 16 ticks, and a 16-tick post delay; wide uses drive 1/16, pre delays over eight ticks, and a three-tick post delay. The quiet fixture generates no spikes under this protocol; it tests state-update and idle event-handling costs rather than active synaptic throughput. The wider case exercises delayed plasticity. Full configurations and topology identities accompany the saved reports.

Candidate selection uses the three randomized profiling rounds with order seed 1729. Five subsequent winner replays form a distinct sequence and are not substituted for the profiling samples. Independent compiled-f32 control results and f64 diagnostic gates are retained. The cache study uses a later source cohort, one full calibration followed by three exact-input hits, and enabled compilation/allocation reuse. Its activation timing includes planning, context checks, execution, hashing and result transport/loading, but excludes Brian frontend lowering. Cold calibration versus later hits therefore measures the combined configured workflow, not an isolated causal effect of the decision cache. Verification covers six archived benchmark-report hashes, selection rules, fingerprints and summaries rather than a repeated hardware campaign or complete raw-array audit.

### 4.6 Compatibility and learning validation

Compatibility evidence is separated into frontend/lowering admission, zero-duration preparation, bounded smoke execution, full-duration execution, and numerical/lifecycle comparison. The staged official-example scanner checks syntax/dependencies, first-run preparation, and bounded execution; it is not a validator of every example's scientific outputs. Full-run tests and explicit numerical checks provide stronger, individually identified evidence. Passing a short smoke run cannot substitute for a scientific or long-duration gate.

Distributed-plasticity tests compare 1/2/4 CPU ranks with the Rust canonical reference using raw state bit patterns. Segment and restart checks include queues, clocks, RNG, refractory state, monitors, and plasticity traces. Native-training validation additionally checks declared VJPs, optimizer state, and forward/carry/recovery behavior. GPU results use their recorded f32 contracts and explicit CPU tolerances. Test totals describe separate frozen source/hardware scopes; overlapping cohorts are not added into a unique count or inherited by later frontend changes. The supplementary matrix distinguishes implemented capabilities, completed hardware tests, and pending combinations.

### 4.7 Published neural-model workflow protocols

The first workflow uses the explicit/general Brian2 NMDA benchmark released by Skaar, Haug and Plesser 15, frozen at 68e6dd970cfc6bab26459fcb8c34ee0f16560d9e. Every excitatory NMDA edge retains nonlinear rise and gating state. The source workload includes recurrent connectivity, 0.5 ms delays, RK4 at dt = 0.1 ms, f64 state and the original 28-field monitoring scope for one second of model time. Network sizes span 2,560–20,480 neurons and 11,796,480–754,974,720 total synapses. The benchmark executes the original explicit formulation, rather than substituting the paper's scientific approximation.

Deterministic validation replaces the unseeded external-input generator with declared identical event tables in both implementations. This diagnostic preserves the recurrent equations, connectivity, delays and update/monitor schedule, but is separate from the unchanged-source performance workload. Same-input gates extend through 10,240 neurons; the 20,480 case retains structural, repeated-output and statistical evidence without a full deterministic gate at that size. The formal Linux CPU cohort pins both backends to CPUs 0–7 with eight workers, one excluded warmup and five measured repetitions per backend, except six at 5,120 neurons. Its compiled-region interval includes native initialization, simulation, monitoring and result dump, and excludes Python preparation, code generation, compilation and public-array backfill. This differs from the original FlyWire simulation/recording interval. A separate thread/MPI control uses the exact same 40 physical CPUs and five repetitions, measuring its stated simulation/recording interval.

The additional CPU cohort uses the contextual-dendritic-gating model repository 14, frozen at dbb77525f2662199544f5a0d3dcc9c18b0e1c853. Two single-neuron workloads exercise nonlinear NMDA and its linear control for 10 s. A separate 100 ms protocol uses the complete Figure-3 network topology, with approximately 1.11 million plastic edges, without claiming validation of its full 45 s scientific protocol.

Both backends execute sequentially on the same remote Linux host with process affinity fixed to logical CPU 190. Campaign order is Cython, Rust, Rust, Cython. Each activation discards one warmup and records five unprofiled samples for a single-neuron variant or three for the network, yielding ten or six measured samples per backend. The primary interval is simulation plus required state recording, excluding construction, export, generation, compilation, result dumps, and warmups. Optional profiles are separate diagnostics. Archived aggregates retain numerical-state gates and timing boundaries; verification covers those records and their arithmetic rather than a new scientific-array audit.

The common-source network retest re-exports the model from the frontend frozen for that measurement cohort with the same frozen topology and named random streams, and validates it using the matching Rust runner. Rust 1.98.1 uses opt-level 3, one codegen unit, panic-abort and the default CPU target; C/C++ dependencies use the existing Zig 0.16.0 toolchain. On the same host and CPU 190, activation order is Cython, cache-disabled Rust, default Rust, default Rust, cache-disabled Rust, Cython, retaining one warmup and three unprofiled samples per activation. Twenty-one state/topology arrays are reread; integer events and topology require exact equality, and floating arrays use rtol 1e-12 and atol 1e-14. The four model exports are identical, and default/cache-disabled native final-state and event dumps are byte-identical. A separate active fixture checks both reduction directions, an offset subgroup, multiple clocks and continued runs against NumPy.

### 4.8 Connected synthetic weak-scaling protocol

The additional study increases network size with host count across 3, 6, 12, 24 and 30 physical EPYC 9454 hosts, using eight populated single-threaded MPI ranks on eight distinct physical cores per host. Each repeated three-host block adds 86 million neurons, arranged in 19 excitatory populations containing 80% of neurons and five inhibitory populations containing 20%. One rank owns each complete population. The three hosts in a block own six excitatory/two inhibitory, six/two, and seven/one populations, respectively. Per-rank neuron and incoming-edge workloads therefore remain approximately constant. The five configurations have 24, 48, 96, 192 and 240 populations and 86, 172, 344, 688 and 860 million neurons.

Every directed population pair has fixed-total uniform random connections with replacement, permitting autapses and multapses. Counts are integer floors of source-count times target-count times 1,000 divided by total neuron count; largest-remainder allocation preserves exactly 1,000N edges. This fixes expected mean indegree rather than the indegree of every neuron. Model seed is 20261007 and projection seeds are 2026100700 + source×P + target in numeric population-creation order, where P is population count. The graph realization changes with size. Whole-population target-owner-local construction preserves source-major connection identities and avoids per-projection count/routing construction collectives.

Euler LIF dynamics use reference-f64, dt = 0.1 ms, membrane and current time constants of 20 and 5 ms, drive 1.05, threshold 1, reset 0 and a 2 ms refractory interval. Excitatory weights are uniform in [0.0027, 0.0033], inhibitory weights in [-0.0132, -0.0108]. Clipped-normal delays have mean 1.5 ms, standard deviation 0.25 ms and bounds 0.1–3 ms. Initial voltage cycles through 1,000 deterministic values. All spikes and voltage/current traces from two neurons per population are retained for 100 ms.

All five configurations use one isolated engine snapshot, with an opt-in neuron ceiling of 860 million and an unchanged default ceiling of one million. Initial-value and IR-byte budgets are four billion values and 64 GiB. Worker cgroup limits are 665 GiB for the first three layouts and 655/660 GiB for the final two (six-excitatory/two-inhibitory versus seven-excitatory/one-inhibitory host layouts), with 768 GiB on the coordinator host; swap is disabled. Each launch requires a fresh availability check with a 64 GiB host-memory reserve at 3–24 hosts and a 48 GiB reserve at 30 hosts, a 128 GiB disk reserve, a 64 GiB per-file ceiling and a 1,800 s execution timeout. Preparation and output auditing have separate 320 GiB/four-core limits and 7,200/1,800 s timeouts. The final worker-admission guard changes only the positive memory-headroom threshold; preparation and audit guards retain a 64 GiB reserve. The hosts are shared resources; the study reports one observation per size without inferred uncertainty intervals.

Each layout first passes a separate small-model gate, using 24,000 neurons and 24 million edges per three-host block and the same population/ownership structure. Complete output is compared with the independent single-process Rust reference, requiring exact spikes and event totals and state/trace differences no larger than 1e-12. Large-run acceptance requires complete output reading, finite state/trace values, refractory and spike-count consistency, neuron/edge/event totals, populated rank and host identities, distinct allowed physical-core coverage, exclusive target-owner construction, successful terminal resource records and cleanup. Percentages use an 86-billion-neuron reference count, rounded from the reported adult human neuron-count estimate 16; they describe count normalization only. Source identities and complete records are provided in Supplements S19 and S20; historical fixed-host capacity cohorts remain separate in S19.

## 5. Results

### 5.1 Semantic conformance and supported behavior

The linked-summed example exposes a cross-population dependency missed by considering only final output. Its regression compares reference, AOT, and Brian NumPy. The implementation also rejects invalid IR, unsupported capabilities, altered plans, and mismatched instance identities. These boundaries make unsupported behavior explicit rather than silently changing execution.

The CPU full-connectome cohort supplies a larger execution comparison. All 80 formal replays, plus warmups, passed same-host Rust/C++ checks with zero maximum difference in the checked arrays. Each run emitted 3,853,674 spikes; Rust counted 462,577,974 weighted-edge deliveries. Across the two hosts, spikes, recorded trajectories, and refractory state agreed, while final voltage differed by at most 2.220446049250313e-16. Exact same-host results and a small cross-host difference are kept distinct.

These results support the tested models and observations, without collapsing the boundaries in Table 1. MPI accepts a narrower model set than generic reference, and WebGPU a narrower set than single-machine Metal/CUDA. Comparisons using shared reference code cannot exclude bugs shared by that runtime or specification.

### 5.2 Distributed capacity and resource use

Partition-local execution reduced memory and time relative to the earlier MPI implementation in the FlyWire EI cohort. At one, two, and four ranks, maximum per-process RSS fell from 403.5, 400.5, and 400.5 MiB to 210.0, 123.0, and 76.5 MiB. Median simulation/recording/gather times fell from 12.1922, 12.8721, and 13.2456 s to 4.2326, 4.4974, and 4.0587 s, respectively (Figure 3). Complete results and events agreed byte-for-byte with native reference in all 33 recorded runs: six warmups, 18 measured resting runs, and nine additional condition checks.

This establishes an implementation improvement and reduced per-rank memory, but little strong-scaling evidence for the optimized version. Four ranks improve its one-rank median by approximately 4.3%, while the four-rank sample range spans 7.54% of the median; two ranks are slower. Four-rank spike exchange and ordering consume approximately 2.92–3.24 s, with 30,000 population/tick exchanges. A separate single construction observation reduced project-generation/compilation peak memory from a historical 7.49 GiB to 663.7 MiB.

The multi-area study demonstrates capacity. A frozen chi = 1.9, seed 1729 configuration with 32 areas, 254 populations, and 8,344 projections completed 100.5 s of model time. It contained 4,129,924 neurons and 24,126,516,728 recurrent synapses on four nodes, with eight ranks per node. Recorded wall time was 39,301.456712 s, approximately 10 h 55 min, and measured CPU use was 196.64248015 core-hours. Independent per-node proxy peaks were approximately 178.5–184.8 billion bytes, below each 256 GiB ceiling. Terminal checks and the subsequent raw-output audit passed.

The original resource record establishes completion of that instance, recording path, and configuration. Its 43.66828524 participating node-hours concern shared hosts, not exclusive allocation or monetary cost. Figure 3 retains this historical cohort rather than combining later runs into a scaling curve.

Subsequent evidence includes Rust seeds 1729, 1750, and 1751, plus full-duration exploratory native NEST references for seeds 1729, 1730, and 1731. The latest seed-1751 raw-output gate and six descriptive estimators are recorded as complete, and the three-Rust-realization descriptive cohort has verified collection receipts. These strengthen repeat-execution and output-analysis evidence at the demonstrated size. They are not independent timed repeats under one matched performance protocol.

Scientific and performance acceptance remain separate. Comparisons retain systematic LvR differences and unresolved historical sampling conventions; a modern V1 spectrum view is not an established match to the published sample. Three exploratory NEST references do not supply prospective equivalence margins, multiplicity handling, or adequate power for a confirmation claim. Different parallel layouts, output policies, and timing scopes also prevent a defensible NEST speed or cost ratio. The implementation therefore reaches a problem-size regime addressed by established distributed simulators, while biological equivalence, competitive efficiency, real-time execution, and scaling to hundreds of nodes remain unestablished.

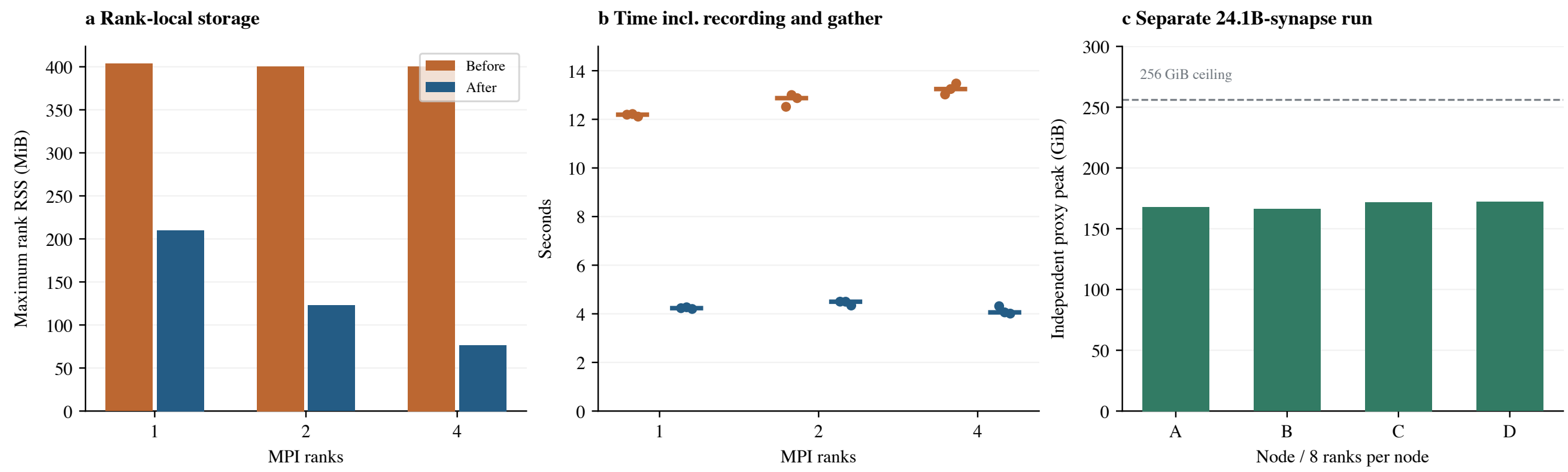

**Figure 3 | Local-storage improvement and scaling are distinct.** Left: maximum single-rank RSS across measured resting runs. Middle: individual simulation/recording/gather samples and medians before and after the change, excluding warmup. One rank uses one node; two/four ranks use two nodes. Right: independent proxy peaks from the separate four-node multi-area capacity run. The cohorts use different snapshots and are not points on one scaling curve.

## 5.3 CPU/GPU performance and memory-constrained execution

On the Linux full-connectome workload, the best measured Rust configuration took 0.839 s at 16 threads, versus 4.555 s for the best measured C++ configuration at eight threads: a 5.43-fold median-time ratio. One-thread values were 4.013 and 7.169 s. Rust's one-to-sixteen-thread improvement was 4.78-fold. The host had two EPYC 9454 processors, but both backends were restricted to a 16-physical-core, single-socket placement. These results do not describe full-machine 96-core performance.

The M1 Ultra host showed a best-measured ratio of 5.05-fold: 0.767 s for Rust at 16 threads and 3.873 s for C++ at one thread. C++ four- and eight-thread observations exceeded the predefined 15% variability threshold and remain descriptive. Selected-configuration native RSS was 400.95 versus 829.05 MiB on M1 Ultra and 317.19 versus 768.29 MiB on Linux. Figure 4 presents thread curves and memory without converting native RSS into frontend-inclusive memory.

The PD14 resource test completed 10 s of model time for 77,169 neurons and 298,880,968 synapses on a 16 GB M3 laptop. Rust's simulation body took 104.009 s and frontend/build/run/load took 125.586 s, with a 6.314 GB native-child peak and a separately measured 0.319 GB frontend peak. Recording was disabled. The C++ attempt, requested for only 0.1 s, exceeded its planned memory budget while materializing connections and parameters before compilation. It was stopped under severe swapping; no kernel OOM kill was observed. This demonstrates preparation and capacity differences without a local C++ simulation-time denominator.

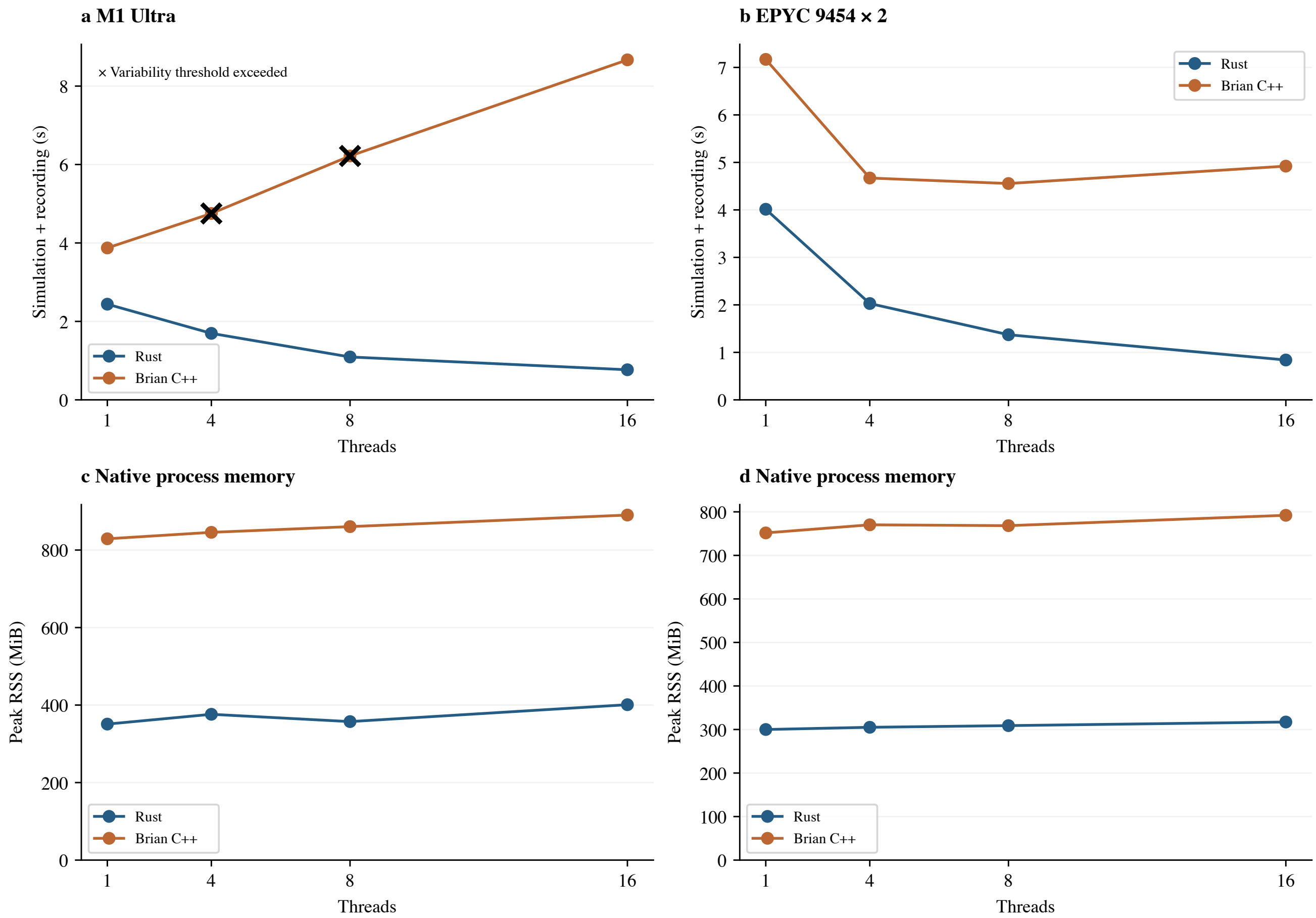

**Figure 4 | Full-connectome CPU measurements.** Simulation/recording medians are transcribed from the retained five-repeat cohort report. Native RSS is separate. M1 Ultra C++ four- and eight-thread points are marked as unstable by the report's criterion. Raw CPU samples are absent from the accompanying evidence package; the figure therefore shows report-derived medians without error bars. The PD14 case is treated as capacity rather than a completed speed comparison.

At 16,384 neurons, delayed-STDP ring replay medians were 67.19 ms for M3 Metal, 34.82 ms for L4 CUDA, and 32.58 ms for A100 CUDA. Corresponding Rust f64 slot-executor medians were 85.73, 216.02, and 181.80 ms on those respective hosts. All four own-GPU policies matched complete compiled f32 control arrays exactly. Prefix/bitset choices did not uniformly improve medians, and this fixture does not establish superiority over optimized multicore CPU simulation.

External comparisons require qualification details. At the larger ring size, the original direct-GeNN adapter failed bootstrap checks on both NVIDIA devices and is excluded from timing. Explicitly labeled barrier variants qualified. Brian2CUDA and corrected Brian2GeNN had much longer complete replays in this fixture, but their adapters also have differing process, file, and load/unload costs. These are not equivalent kernel-speed ratios. External variants and failed qualifications remain in the supplement.

Recurrent CUBA gives a counterexample. For 4,096 neurons, 131,072 connections and 2,048 ticks, native CUDA took 116.79 ms on L4 and 137.13 ms on A100. Direct GeNN with the declared Brian-Euler adapter took 52.79 and 52.11 ms; Rust CPU f64 took 58.92 and 48.46 ms. Native CUDA remained slower than both despite passing matched numerical gates. The f32/f64 diagnostic failed at this scale and remains distinct from the passed f32 gate. Figure 5 points to workload, dispatch, transfers, and lifecycle costs rather than a platform-wide ranking.

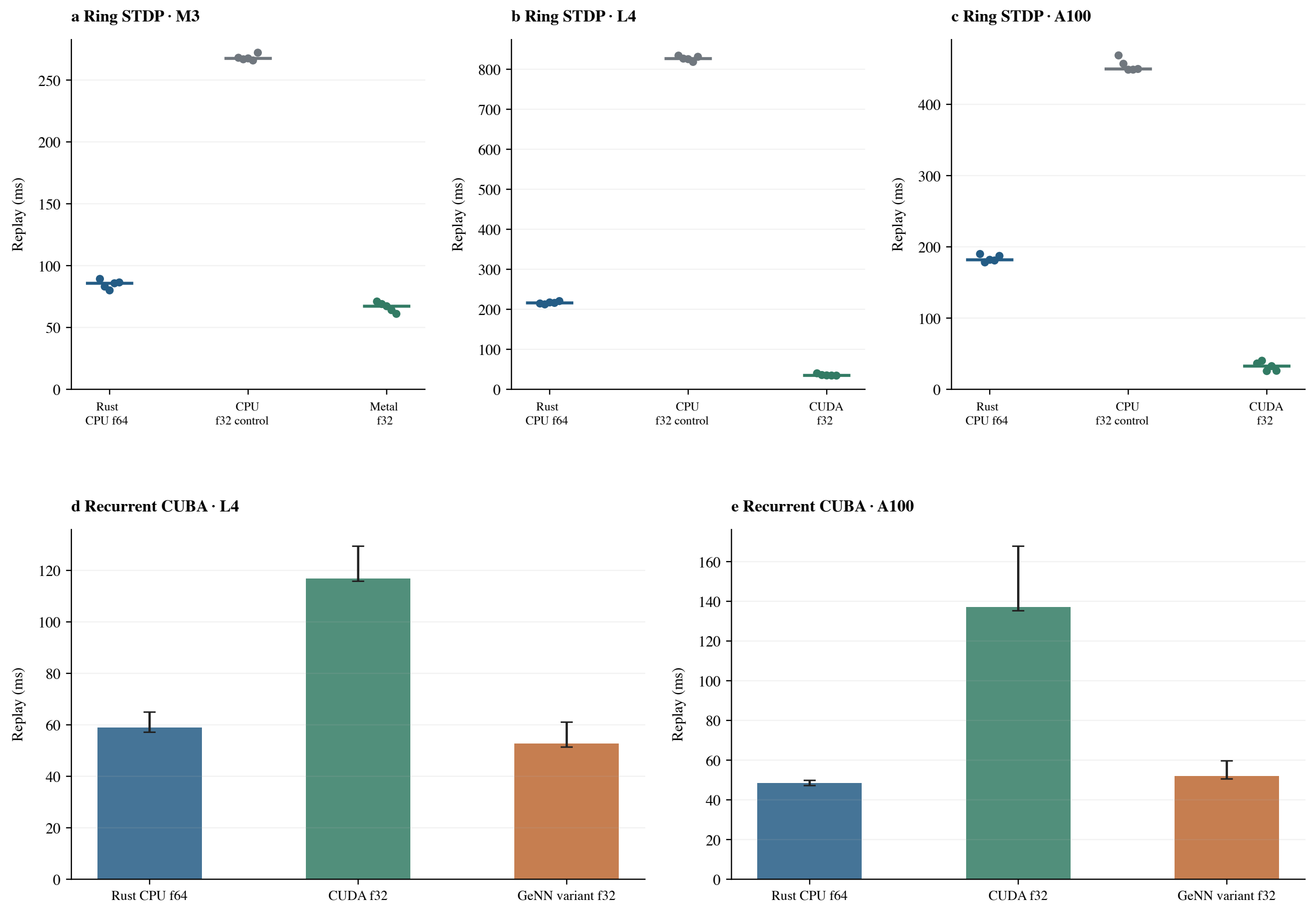

**Figure 5 | GPU gains depend on workload and lifecycle.** Top: five recorded replay samples and medians for the 16,384-neuron ring, separated by host. Bottom: recurrent CUBA medians and reported min–max ranges. Rust is f64; GPU and expression controls are f32. Direct GeNN in the lower panels uses the disclosed Brian-Euler adapter. All intervals cover reset through complete host arrays, not isolated kernels.

## 5.4 Long-duration recording and recovery

The assembly variant exercises mutable synapses, conductance dynamics, homeostatic operations, and long execution. Independent repeated confirmations showed lower compiled simulation/recording medians for selected Rust configurations on M1 Ultra and Linux. Gains were smaller than for non-plastic FlyWire: full 0–11 s intervals gave ratios of 1.2697 and 1.0829. Linux used 32 Rust workers and four C++ workers, and its one-worker selection result favored C++. These observations do not establish better per-core efficiency.

In an earlier one-worker memory cohort at 1,000 s, native RSS was 1.082 GB for full Rust recording, 0.290 GB for a one-second rolling window, and 2.327 GB for C++. Corresponding summed process-tree peaks were 4.961, 4.074, and 2.767 GB. Frontend and data representation therefore outweighed native savings in this configuration (Figure 6). These are single observations from a version preceding the final parallel edge-runner change, not the same snapshot as final performance measurements.

Three fresh processes recovered the training checkpoint and replayed all 1,000 spontaneous seconds, matching 41 final fields and all 11 recorded trajectory segments exactly. This establishes the tested process-restart behavior. A separate short restore/continue observation took 52.685 s for Rust and 2.329 s for NumPy including preparation and checkpoint I/O. Fast compiled execution consequently does not imply fast recovery preparation. The application also showed spontaneous-rate drift and does not establish stationary attractor dynamics or full reproduction of the original model.

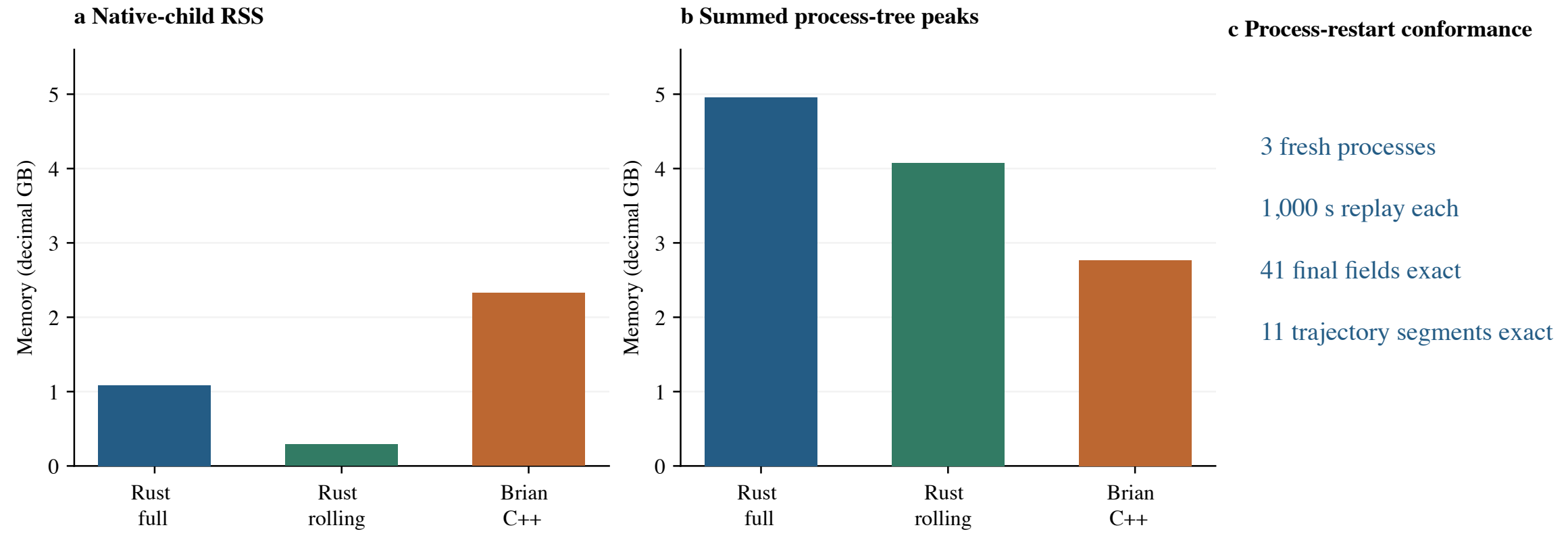

**Figure 6 | Recording memory and restart conformance.** Bars show a 1,000 s single-observation cohort in decimal GB. Native-child RSS and summed process-tree peaks are distinct. The recovery summary concerns three fresh-process replays and is a correctness result, not a durability benchmark. A one-second rolling window does not eliminate memory growth elsewhere.

### 5.5 Browser portability and interactive execution

Generic WASM provides a workflow from validated model bundle to local simulation and export. Retained checks cover native/WASM comparison, batch invariance, integrity, cancellation, and bounded equation authoring. Sharing reference runtime code tests portability and host behavior while leaving shared bugs possible. WebGPU remains an experimental independent-cell profile, with observed f32/f64 differences preserved.

Neural Lab exposes this workflow through an interactive browser workbench. Users select a model and execution profile, edit parameters, run an experiment, inspect spike and state recordings, and export the executed configuration with its validated bundle. Figure 7b shows the actual interface after a local WASM/f64 run of 160 independent adaptive leaky integrate-and-fire neurons for 600 ms, with a 0.1 ms timestep and seed 42. The run produced 3,706 spikes, corresponding to 38.60 Hz per neuron. The workspace connects model equations and controls to a spike raster, population-rate curve, firing-rate distribution, and selected-neuron voltage and adaptation traces. This single run illustrates the implemented interaction and visible output; it is separate from the conformance cohorts and the model-specific AOT application below.

Model-specific WASM AOT executed the full graph of 139,255 neurons and 15,091,983 weighted edges. Thirteen fresh simulations covered fixed inputs for digits zero through nine and an A/B/A sequence. Input/activity hashes and predictions matched the CPU oracle; maximum absolute score difference was 7.413514246934483e-14. These checks establish fixed-input and repeated-input behavior, not classification accuracy on an independent test set. No new 10,000-image evaluation is reported.

WASM linear memory was 559,742,976 bytes, approximately 533.8 MiB; model resources were approximately 54 MiB. This excludes JavaScript, rendering, and transient loading copies. Browser checks stopped the static server, reloaded from cache, and continued local execution. This demonstrates the tested offline path without guaranteeing persistent caching or mobile-device memory fitness. Figure 7a separates generic execution, WebGPU, and model-specific compilation; the screenshot in Figure 7b uses only the generic WASM path.

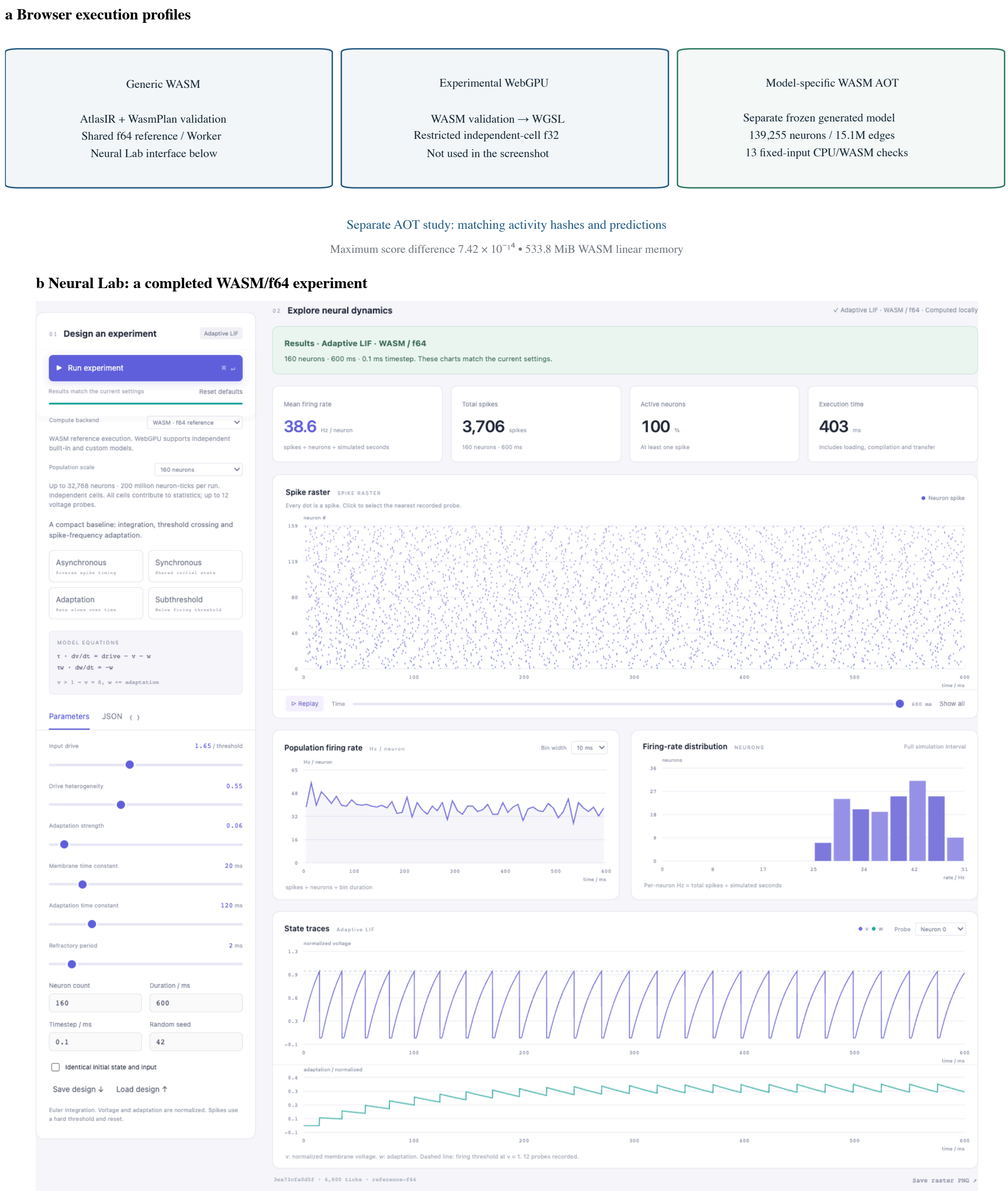

**Figure 7 | Browser execution profiles and an interactive simulation workbench.** (a) Generic bundles use the validated shared reference runtime. WebGPU supports a restricted f32 independent-cell subset. Separate AOT uses a frozen generated model and memory-only host; its 13 fixed-input checks concern CPU/WASM agreement, with additional offline workflow checks. (b) Unmodified Neural Lab workspace screenshot after a local WASM/f64 adaptive LIF run: 160 independent neurons, 600 ms, dt = 0.1 ms, seed 42, and 3,706 spikes. Controls, equations, output summaries, raster, rates, and state traces belong to the same completed run. The displayed 403 ms is one UI observation including loading, compilation and transfer, not a benchmark. This screenshot does not demonstrate WebGPU or full-connectome AOT execution. Capture provenance and the exported experiment accompany the supplement.

### 5.6 Strategy selection, calibration cost and decision reuse

The planner connects the semantic counterexample in Figure 2 to concrete implementation choices: completed cross-population reads prevent final-only evaluation, while unsafe fixed-phase schedules select a supported canonical path. Plan records expose the selected implementation and its reasons. These observations establish an executable planning boundary; the aggregate CPU speedups above do not isolate the contribution of planning from the kernels, layouts and runtime policies it selects.

The GPU calibration study selected ordered bitmaps for the quiet NVIDIA cases and prefix plus bitmaps for the wide NVIDIA cases (Table 3). Within the selection rounds, L4 wide decreased from 122.26 to 79.86 ms, and A100 wide from 154.33 to 91.75 ms, reductions of 34.7% and 40.6%. M3 retained baseline in both cases because no alternative cleared the improvement and separated-range conditions. In M3 wide, baseline profiling samples ranged from approximately 359 to 1,845 ms. A lower candidate median alone was therefore insufficient to select it. Baseline retention does not establish baseline optimality.

**Table 3 | GPU policy selection and calibration cost.** All cases have 4,096 neurons and 1,024 ticks. Profiling medians use the three selection samples; later medians use five separate winner replays. Total calibration includes preparation and selection overhead. Values are rounded from retained machine-readable reports; no timing intervals are reconstructed.

| Host / case | Selected policy | Baseline profiling (ms) | Selected profiling (ms) | Later winner replay (ms) | Total calibration (s) |
|---|---|---:|---:|---:|---:|
| M3 quiet | Baseline | 79.63 | 79.63 | 63.60 | 4.31 |
| M3 wide | Baseline | 1,313.67 | 1,313.67 | 323.75 | 16.87 |
| L4 quiet | Bitset | 24.08 | 17.29 | 17.40 | 15.99 |
| L4 wide | Prefix + bitset | 122.26 | 79.86 | 81.00 | 24.36 |
| A100 quiet | Bitset | 30.90 | 20.84 | 19.86 | 14.91 |
| A100 wide | Prefix + bitset | 154.33 | 91.75 | 100.19 | 23.99 |

All selected results and the five later replays passed the retained f32 gates and matched the complete compiled-f32 control. The quiet f64 diagnostic passed; the wide f64 diagnostic failed and remains a distinct limit. Calibration took 4.31–24.36 s, substantially exceeding a single replay. The later M3 timing shift also demonstrates why separate measurement sequences must not be treated as interchangeable estimates. These results support verified selection among measured candidates, without establishing end-to-end acceleration over one manually configured run.

In the separate decision-cache cohort, three exact-input hits per case skipped calibration while preserving complete-result checks. Across the six cases, initial calibration plus transport took 3.707–26.255 s and median hit activations took 0.560–4.203 s. Policies differed from the earlier cohort in some cases, including A100 quiet retaining baseline. All recorded hit outputs passed f32 checks, while the wide f64 diagnostic still failed. Compilation and buffer reuse were enabled and contribute to this configured workflow; the difference is not attributed solely to caching a decision. Per-case results and the scope of the retained lifecycle tests appear in Supplementary Section S9. Changing the state or activity can require calibration again, and a cache hit does not guarantee continued performance optimality.

## 5.7 Stateful MPI and native-training functionality

The distributed-plasticity delivery cohort records 30 passed tests: 19 for training/continuation/recovery and 11 for compact queues. CPU Pair/Triplet tests use 1/2/4 ranks and raw-bit comparison to the canonical reference. Fresh-process MPI examples preserve time, weights, spikes, and pending state; local CPU/Metal examples exercise the explicitly limited neuron-update offload. Small probes across six LIF application families exercise training, freezing, fixed-candidate reconnection, and restored replay. These are functionality and determinism results for the recorded source version, not production-scale training or fault-tolerance benchmarks.

A separate first native CPU/Metal cohort records 20 passed tests covering surrogate BPTT, declared reset/detach choices, masks, truncated windows, and fresh-process optimizer replay. Later frozen training cohorts broaden ordered dynamics, stochastic programs, and local MPI tests. One CPU/Metal pipeline snapshot records 3,194 passed tests and 1,181 explicit skips across 53 modules; the later indexed-endpoint snapshot has separate CPU and Metal/local-MPI acceptance, while the subsequent external-input ownership snapshot has CPU/local-MPI acceptance only. Neither result is inherited from that earlier pipeline run. A historical static CUDA training stage was tested on a single L4; newer dynamic CUDA results are recorded in the separate qualification below. Cross-host and multi-physical-GPU training remain unqualified. Supplement S12 records the scopes. No complete public-dataset accuracy or fair training-throughput ranking is inferred from these tests.

The Modal L4 qualification passed all 708 previously skipped CUDA cases from the three captured training increments and one compiled-library ABI regression (709 passed, zero failures or skips). The scope covers 15 training modules, including single-process execution and two MPI ranks sharing one GPU. Four missing C-linkage declarations were repaired before this run; the initial failures and complete rerun records are retained. These are functional and numerical checks, not training-performance measurements.

## 5.8 Validation on published neural-model workflows

### 5.8.1 Explicit NMDA: state-heavy execution and distribution

The explicit NMDA workload exercises clock-driven nonlinear synaptic ODEs, delayed pathways, postsynaptic summed reductions, derived monitors and population rates through the generic Brian2 → AtlasIR → execution-plan route. General improvements parallelize eligible per-edge RK4 state work and omit unused event-routing structures for purely clock-driven/summed connections. These changes are selected by model dependencies, rather than by recognizing a paper or model name.

At 10,240 neurons, the same-event diagnostic passes its exact rate/time/topology requirements and strict f64 state bounds. Maximum voltage and NMDA-gate differences are $8.33 \times 10^{-17}$ V and $3.72 \times 10^{-15}$. Independent-input-stream screens that failed on final per-edge state at 5,120 neurons remain recorded; the same-event diagnostic defines a different, explicitly declared comparison and does not erase those failures. At 20,480 neurons, the full 754,974,720-synapse configuration completes with all 28 finite public fields and repeated Rust output identities, while deterministic cross-engine validation remains limited to the smaller scales.

**Table 4 | Published explicit NMDA workload, matched eight-core CPU cohort.** Both backends run one second at dt = 0.1 ms with RK4/f64, original delays and monitoring. Times are compiled-region medians in seconds, including native initialization and dump as defined in Section 4.7. Ratios compare medians within each frozen cohort. Reported native sampled-RSS increases are distinct from total process memory.

| Neurons | Total synapses | Samples per backend | Brian2 C++ (s) | Atlas CPU (s) | Brian2 / Atlas | Reported RSS increase |
|---|---|---|---|---|---|---|
| 2,560 | 11,796,480 | 5 | 69.942 | 61.998 | 1.13 | 5.5% |
| 5,120 | 47,185,920 | 6 | 289.100 | 194.660 | 1.49 | 7.3% |
| 10,240 | 188,743,680 | 5 | 1,256.384 | 598.964 | 2.10 | 7.9% |
| 20,480 | 754,974,720 | 5 | 4,443.328 | 2,271.710 | 1.96 | 8.3% |

The tested gains coexist with approximately 5–8% higher sampled native RSS. Compiler targeting and parallel coverage are part of the measured implementation; the result does not isolate an IR-only speedup. On a separate dual-socket host at 10,240 neurons, the same-40-core control has simulation/recording medians of 257.774 s for 40 shared-memory workers and 153.675 s for 40 MPI ranks, a 1.68-fold ratio (Figure 8). Fixed-total-rank placement experiments across machines expose saturation under per-timestep spike exchange. They demonstrate placement behavior, not increasing-resource strong scaling. Accelerator evidence uses f32 gates and is not divided by the primary f64 CPU timing.

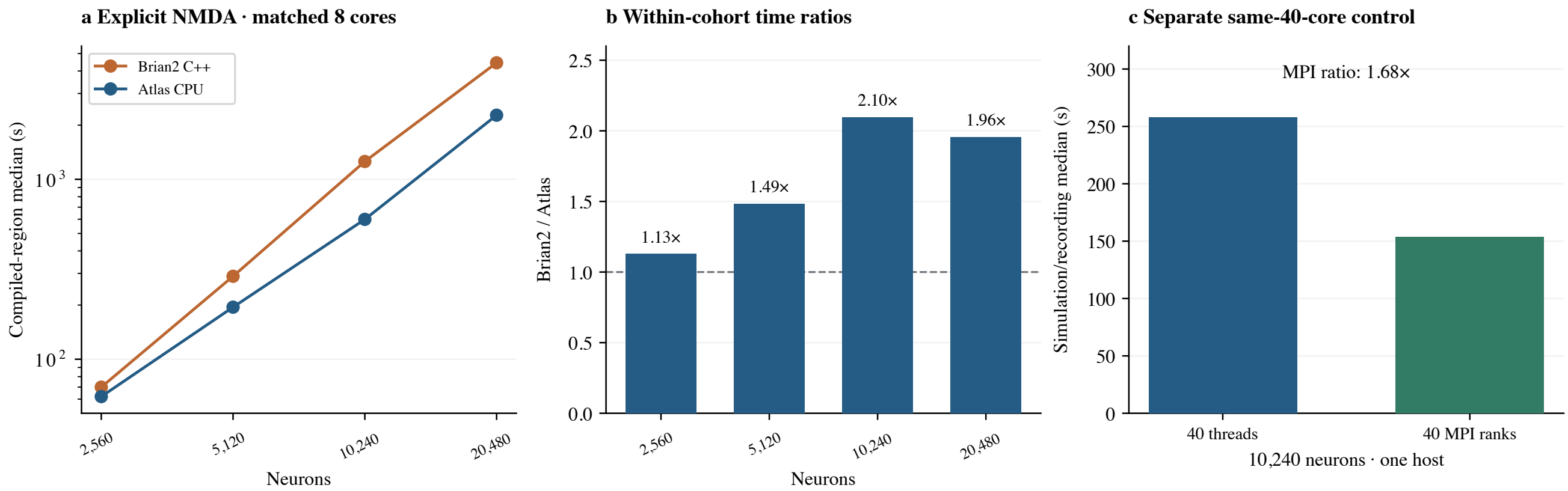

**Figure 8 | NMDA CPU scale and same-core MPI control.** (a) Compiled-region medians for the four matched eight-core cohorts in Table 4. (b) Their Brian2/Atlas median ratios. (c) A separate 10,240-neuron cohort compares thread and MPI simulation/recording medians on the same 40 physical CPUs. These timing intervals and cohorts remain distinct; no uncertainty intervals are reconstructed. Exact-input state validation extends through 10,240 neurons, while the 20,480 result establishes the separately audited capacity/output scope. The associated exact/approximate NEST decision-network reproduction is a scientific-model comparison, not an Atlas backend result, and is described separately in the supplement.

### 5.8.2 Contextual dendritic gating: specified scientific results and paired engine checks

The second workflow combines dendritic dynamics, NMDA conductances, voltage-based plasticity, contextual inhibition, stochastic inputs and normalization. Reproduction of the source model in its locked Brian environment and paired Atlas validation answer different questions. For the published Fig. 2 and S1 parameter scans, all ten seeds and 32,000 grid cells per figure were rerun. Ensemble-mean correlations to published surfaces are 0.999604 and 0.999513, with weight-change sign differences of 2.375% and 0.09375%, respectively. These source-environment reproduction records are not 32,000 Atlas/Brian paired executions.

Separate 200 ms Atlas/Cython gates record dendritic trajectories and weights under controlled matching inputs/noise. The nonlinear and linear fixtures have one and four exact soma spikes, respectively, with maximum dendritic trajectory differences 3.05 × 10^-16 and 2.64 × 10^-16. Both pass rtol 1e-12 and atol 1e-14. Paper-duration 10 s gates and formal timing campaigns retain their narrower final-weight scope.

The two 10 s single-neuron variants pass their archived final-weight gates at rtol 1e-12 and atol 1e-14 across the four campaign activations. Their median Cython/Rust ratios are 12.49 and 14.75 (Table 5); these archived measurements retain their original implementation identities.

On the common-source snapshot, the regenerated network has Cython and default Rust medians of 9.796826 and 7.631757 s, giving a 1.28-fold Cython/Rust ratio. A same-source cache-policy ablation reduces Rust runtime by 34.0% with endpoint-expression reuse enabled (S13). Separate node diagnostics identify repeated NMDA exponentials in three summed updates; legal endpoint reuse reduces their source-level evaluation count from approximately 1.113 billion to 7.2 million per 100 ms while preserving edge accumulation order. All 21 paired state/topology checks pass, and the complete native state/event dumps are exact across cache policies. The 100 ms protocol does not qualify the full 45 s scientific workflow or its speed.

**Table 5 | Additional dendritic CPU workloads.** All times are reported steady-state simulation/required-recording medians in seconds. Ratios are Cython divided by Rust. Samples are measured observations per backend; no confidence intervals or end-to-end ratios are inferred. The first two rows are archived single-neuron cohorts with their original implementation identities; the final row uses the common-source snapshot with production defaults. The same-source cache-policy ablation is reported in S13.

| Workload | Model time | Samples per backend | Brian2 Cython (s) | Rust AOT (s) | Cython / Rust |
|---|---|---|---|---|---|
| Single neuron, nonlinear NMDA | 10 s | 10 | 10.346926 | 0.828325 | 12.49 |
| Single neuron, linear NMDA | 10 s | 10 | 10.273518 | 0.696412 | 14.75 |
| Figure-3 complete topology, common source, default | 100 ms | 6 | 9.796826 | 7.631757 | 1.28 |

These results extend the measured workload range beyond uniform LIF without making the new Cython baseline interchangeable with the earlier C++ comparisons. They also show why an execution-core replacement should be assessed by model structure and timing stage. The remaining figure families retain pending or failed scientific gates, including ensemble recall and association comparisons; Supplement S14 preserves the detailed matrix. Cache-based redraws, source-model reruns, paired engine conformance and matched performance remain separate evidence levels. These admitted results therefore demonstrate specified published workflows rather than complete reproduction of the upstream study.

### 5.9 Heterogeneous CPU/GPU MPI: implementation and qualification

The simulation planner admits an explicit backend per rank, preserving ownership, canonical host-event order and global RNG identities under the selected numerical profile. The retained local Apple Silicon delivery contains 25 passed mixed-GPU/inventory checks. Together with CPU and planning regressions, that delivery has 118 passed and 12 environment-skipped tests; these counts are not pooled with subsequent qualification stages. Its two-rank CPU/Metal example records 82 spikes and successful GPU dispatch counts [0, 32], confirming that the CPU rank performs no GPU dispatch while the Metal rank executes the offloaded update nodes. Tests cover rank order, population assignment, empty owners, delays, refractory state, controlled numerical expectations and failure propagation.

CUDA simulation source generation and a CUDA Runtime adapter are implemented, but this mixed-MPI simulation delivery has no NVIDIA compilation/execution qualification or cross-host heterogeneous result. Separate historical native-training validation does contain real CUDA MPI execution: on one NVIDIA L4, 31 CUDA MPI tests pass within a 100-passed/64-skipped cohort. Those ranks share logical device zero on one host, and execute the separate training-plan family. They do not qualify the AtlasIR simulation-offload adapter, multiple physical GPUs, or simultaneous Metal/CUDA ranks. Supplement S17 gives the evidence matrix and retained record identities.

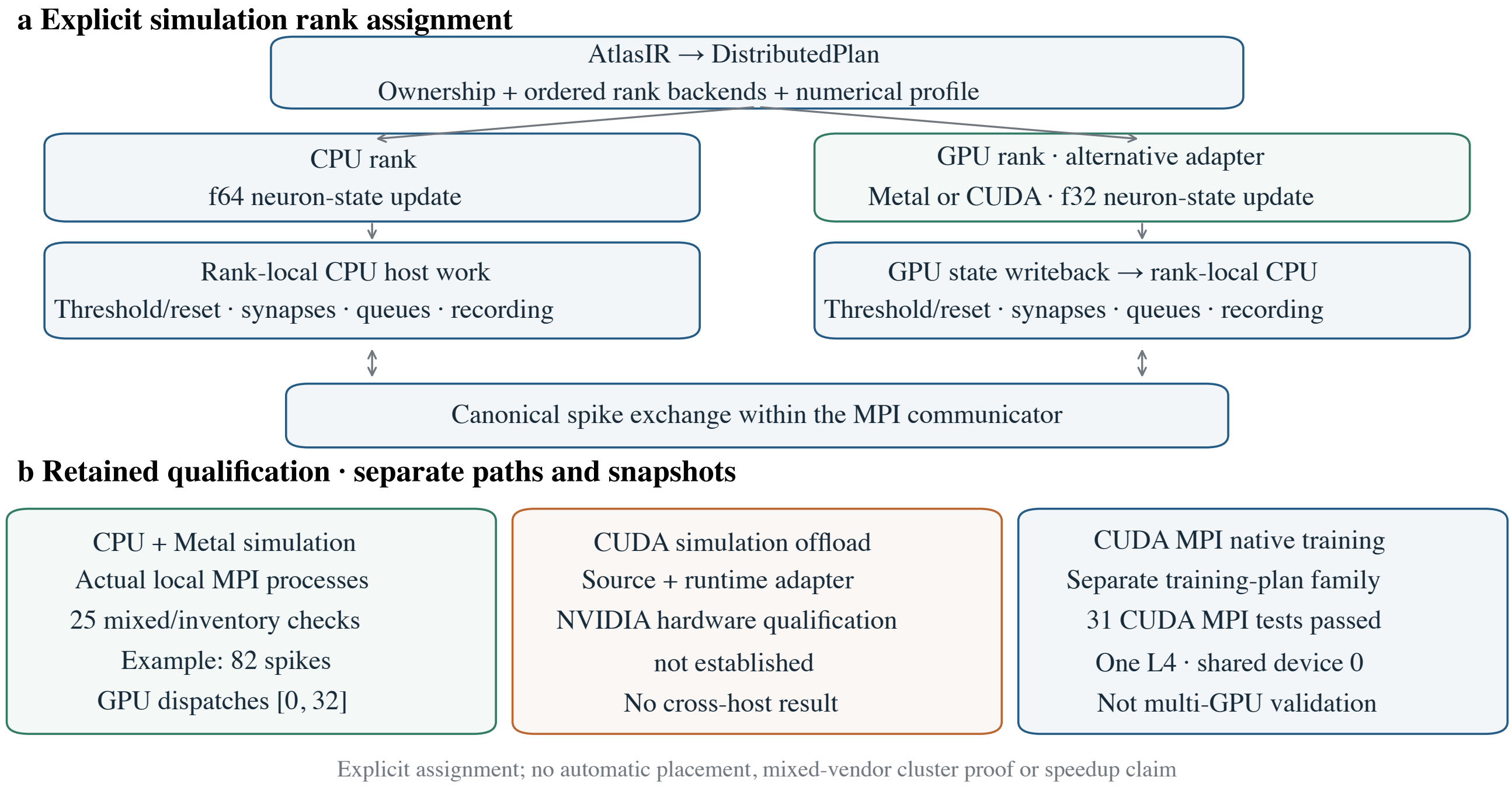

**Figure 9 | Heterogeneous rank assignment and evidence boundaries.** (a) A simulation DistributedPlan assigns CPU or GPU neuron-update execution per rank. The GPU box denotes alternative Metal/CUDA adapters, not a verified combination of both vendors in one communicator. State returns to the host before threshold/reset, synaptic work and ordered spike exchange. (b) The local CPU+Metal simulation delivery, unqualified CUDA simulation-offload path and separately accepted one-L4 CUDA MPI training cohort are distinct evidence scopes. No mixed-mode speedup, automatic device discovery/load balancing, cross-host mixed-vendor compatibility or multi-GPU scaling is inferred.

## 5.10 Large connected synthetic networks and weak scaling

The five configurations completed 100 ms of model time, reaching 860 million neurons and 860 billion connections on 30 hosts and 240 physical worker cores. Counts correspond to 0.1%, 0.2%, 0.4%, 0.8% and 1% of the 86-billion reference, using the same engine snapshot and the protocol in Section 4.8. All layouts passed their small-model numerical gates and complete large-output audits, with successful resource exits and cleanup. These are count-normalized synthetic recurrent E/I networks, not anatomical or functional fractions of a human brain.

**Table 6 | Accepted synthetic weak-scaling observations.** All cases use reference-f64, dt = 0.1 ms, 100 ms model time and exactly 1,000N connections. Eight physical worker cores are used per host. Preparation includes export, code generation, compilation and input hashing. Launch-to-exit excludes preparation, deployment and the independent post-run audit. Stage maxima need not occur on the same rank. Each configuration has one timing observation.

| Reference count | Hosts / ranks | Neurons (M) | Preparation (s) | Launch (s) | Simulation max (s) | Host peak max (GiB) |
|---|---|---|---|---|---|---|
| 0.1% | 3 / 24 | 86 | 240.848 | 605.101 | 236.931 | 560.681 |
| 0.2% | 6 / 48 | 172 | 495.473 | 663.992 | 297.882 | 574.193 |
| 0.4% | 12 / 96 | 344 | 1067.713 | 793.267 | 412.833 | 602.212 |
| 0.8% | 24 / 192 | 688 | 2984.819 | 1049.728 | 612.796 | 658.590 |
| 1% | 30 / 240 | 860 | 4688.101 | 1142.852 | 666.834 | 689.135 |

From three to 30 hosts, maximum local simulation time changed from 236.931 to 666.834 s and launch-to-exit from 605.101 to 1142.852 s. Their descriptive weak efficiencies, baseline time divided by measured time, are 0.355 and 0.529, respectively. Centralized preparation also increased from 240.848 to 4688.101 s and is excluded from both efficiencies. Initialization, communication and collection remain separately recorded. Single observations on shared hosts do not establish a precise scaling law; strong scaling with an immutable model and repeated timings is a separate qualification.

The endpoint retained 1,087,452,431 spikes, counted 1,062,437,152,208 delivered synaptic events and produced 46,298,756,762 output bytes. Maximum coordinator-host and other-host worker cgroup peaks were 689.135 and 656.436 GiB, respectively, under the 768 GiB coordinator and 655/660 GiB worker limits; the maximum other-host peak occurred on a host with the 660 GiB limit. Cgroup peaks include charged cache. Host peaks occur at different times and must not be summed as a simultaneous cluster peak. The per-layout numerical gates passed 243, 483, 963, 1,923 and 2,403 checks with exact spikes/events and zero checked state/trace difference; these are small-model comparisons rather than independent reference executions of the large graphs.

Useful local neuron and incoming-edge work remains approximately constant, but global-sized source CSR offsets and spike-receive buffers increase per-rank memory with network size. Ordered spike exchange and result collection also retain global dependencies. The measured capacity therefore does not imply constant memory, constant wall time, or full 86-billion-neuron feasibility through machine addition alone. Earlier 86M/128M/256M fixed-host capacity observations and the smaller 8.6M one-second run use different partitions or snapshots and remain separate in Supplement S19.

Supplement S21 provides a source-derived conditional resource analysis for 86 billion neurons and 86 trillion connections at the same expected indegree. Preserving tested local work maps to 3,000 hosts and 24,000 physical worker cores, conditional on qualified changes to indices, metadata, communication and recording; this is not the current implementation's host requirement or a runtime prediction. Unchanged dense source CSR alone would occupy approximately 641 GiB per rank. The existing generator already splits spike batches before the signed-32-bit count limit, bounding the receive-element payload below 16 GiB per rank; global dissemination, dense projection-by-rank reports and centralized recording remain separate scaling costs. The analysis identifies conditions for further extension without establishing full-scale feasibility.

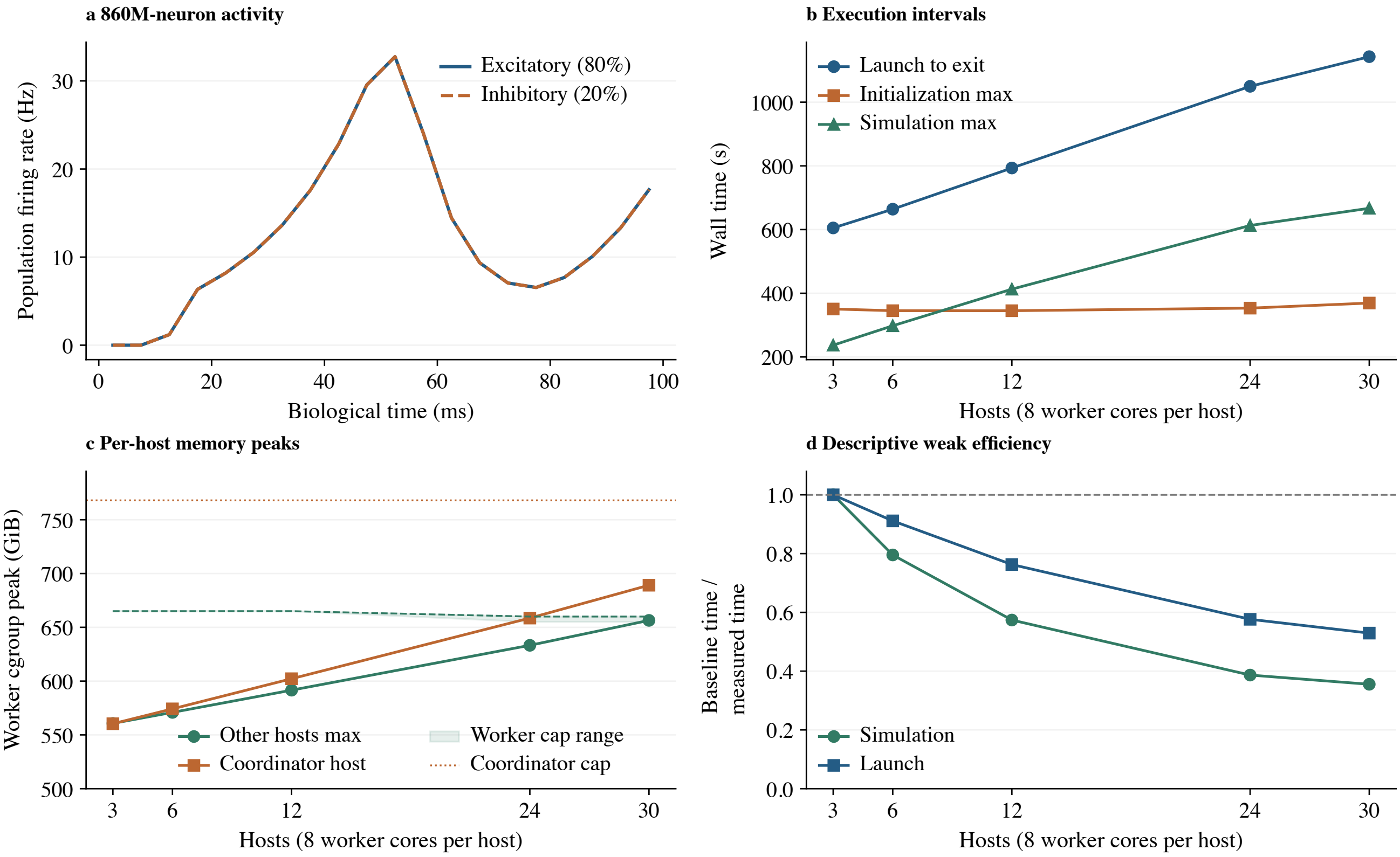

**Figure 10 | Connected synthetic weak scaling with complete-output acceptance.** (a) Excitatory and inhibitory population rates in the 860-million-neuron run, displayed in 5 ms bins. (b) Maximum local initialization/simulation intervals and measured launch-to-exit across the five layouts. Stage maxima may occur on different ranks and must not be summed. (c) Coordinator-host and maximum other-host worker cgroup peaks, including charged cache, with their respective limits. (d) Descriptive weak efficiency relative to three hosts, shown separately for simulation and launch. The workload family has eight physical worker cores per host, 1,000 expected incoming edges per neuron and 100 ms model time. One observation per size is shown, without inferred uncertainty or a fitted scaling law. Graph realizations change with size.

# 6. Related work and discussion

## 6.1 Relationship to existing systems

Brian2's equation-oriented code generation is the retained modeling foundation 1. GeNN provides an independent generated execution system 2, while Brian2GeNN and Brian2CUDA already connect Brian models to GPUs 3, 4. Browser deployment predates this work through Brian2Wasm 12. Frontend reuse, code generation, GPU support, and browser delivery are therefore individually established techniques. Our contribution is the specified AtlasIR boundary and its independently checked, dependency-constrained architecture, together with the implemented ownership, preparation and distributed mechanisms and their measured behavior.

Representations also have important precedents. NIR addresses interoperability through computational primitives for brain-inspired computing 7. SNN-MLIR develops a compilation path from NIR through an MLIR dialect to C 8. AtlasIR focuses on discrete simulation semantics: update statements, canonical scheduling, events, effects, and run identity. This describes our design focus rather than asserting that other representations cannot encode related behavior or that introducing an IR is itself new.

NEST scaling and the MPI–GPU multi-area study provide direct context for distribution, communication and scale 5, 6. Our 24.1-billion-synapse run is an implementation capability result, not a first-of-kind scale claim. Competitive efficiency requires matched models, numerical conventions, output, allocations, and intervals. The current data establish a distributed path, repeated capacity demonstrations, and bounded stateful continuation while exposing communication limits in the smaller cohort. Native training extends the execution mechanisms, but is not part of a demonstrated supercomputer-scale learning comparison.

### 6.2 Consequences of an explicit contract

The contract makes decisions inspectable before execution. A linked consumer can prevent a final-only reduction. Target ownership preserves event ordering while moving storage. A backend can reject unsupported functions or precision choices without substituting a different computation. A browser validates a bundle independently of its exporting Python process. Execution plans connect these constraints to selected layouts and dispatch paths, while explain records make a fallback inspectable. GPU calibration adds measured choice within this constrained space and publishes only a verified final result. These are concrete consequences of representing the relevant semantics at the frontend–engine boundary.

Resource benefits arise at distinct stages. Recipes change preparation and feasible size. CPU ownership affects compiled throughput. GPU persistence and dispatch affect replay. MPI partitioning changes local storage. Recording can dominate total memory despite efficient native state. Keeping these stages separate provides a more useful account than an aggregate speedup across machines and models.

### 6.3 A common workflow across execution targets

The integration objective is to let users keep an admitted scientific model as computational requirements change. Brian2CUDA and Brian2GeNN already reuse Brian's modeling frontend, and Brian2Wasm already supports browser delivery 3, 4, 12. Their existence makes frontend reuse an established convenience. brian2-atlas adds a common downstream contract: one lowering boundary defines state, schedule, events and identity; target plans expose capability decisions; native results return through the retained state/monitor interfaces; and browser artifacts carry verifiable model and plan identities.

For native targets, backend and rank selection are configuration choices within the same Device integration. The browser has a separate artifact and delivery step, and numerical mode remains explicit. Hardware compilers, memory limits, admitted subsets and result/continuation constraints still depend on the target. This common architecture is intended to reduce the number of model-specific adapters and independent execution conventions a user must coordinate, without promising that every script needs only a backend-name change. The published workflows demonstrate migration of demanding model declarations and observations; a controlled usability study would be needed to quantify saved user effort or claim a minimum burden.

### 6.4 Limitations and threats to validity

Compatibility subsets remain explicit, and substantial frontend functionality is retained from Brian2. Shared specification, lowering and reference code introduce common failure modes. Validation and regression tests are not formal compiler proofs. Precision, operation order and library behavior can change recurrent trajectories; a few matching aggregate statistics do not establish equivalence between random streams.

Performance uses bounded searches and heterogeneous development environments. Some hosts were shared or interactive; the Mac CPU cohort includes observations above its variability threshold. Some tables are transcribed from reports whose full raw archives are not yet consolidated into this package. GPU results depend on lifecycle and activity. The original MPI resource cohort has one seed; the later descriptive cohort has three Rust realizations on four nodes. Neither cohort supplies a matched NEST performance experiment, and the smaller study has substantial collective overhead. Neither supports extrapolation to supercomputer-scale efficiency.

CPU work estimates are heuristics, and the GPU tuner evaluates a small fixed policy set using three samples. It does not search backend placement, all dispatch modes or memory-optimal layouts. Calibration overhead, simultaneous candidate storage and changed-input misses limit its practical value; the exact-input cache does not learn transferable performance decisions. The two selection workloads and short measurement sequences cannot establish global optimality or general amortization. An isolated ablation of planning and of decision caching remains necessary to attribute their individual performance contribution.

The maintained repository integrates the execution branches at a committed source revision with scoped acceptance records; this does not establish joint qualification of every feature combination. Historical benchmarks, current implementation inspection, and later frozen regression cohorts have separate identities. Newly implemented endpoint or input semantics require their own target-specific validation rather than inheriting an earlier GPU pass. Browser portability does not make every native storage representation portable, and mixed-rank updates do not establish full distributed GPU execution. Published models and connectome data serve as demanding workloads; biological conclusions require validation beyond this systems study.

The synthetic weak-scaling observations through 860 million neurons each cover 100 ms and one run, using an isolated snapshot with explicitly extended neuron limits. They do not reconstruct human anatomy or establish maximum capacity. This study was self-funded, and execution at the full 86-billion-neuron scale was not evaluated within the available computational resources. Feasibility at that scale remains unvalidated: dense source CSR offsets depend on global network size, projection metadata/reporting grows with populations and ranks, and recording/collection concentrates data on rank zero. The generator already splits consecutive spike producers before the signed-32-bit MPI capacity limit; qualification of batched communication at substantially larger rank counts remains necessary. Frontend count and initial-value limits, IR/output sizes, communication and recording require further evaluation. Supplement S21 separates these implementation costs from a conditional allocation preserving local work. Adding machines alone is not demonstrated to suffice.

Boundary checkpoints currently gather full owner state at the coordinator and prohibit rank, device, ABI, or source migration. Procedural fixed-total capacity and mutable explicit/binary training therefore have different lifecycle contracts. Native-training MPI replicates model/input/tape storage, and truncating the adjoint does not imply bounded streaming tape. Surrogate and stochastic gradient definitions restrict what numerical agreement means. Cross-host and multi-physical-GPU training scalability remains unverified. External training evaluations also contain incomplete qualification and differing executable architectures, so they do not support an architecture-independent advantage over other learning frameworks.

---

## 7. Conclusion

We presented brian2-atlas, a neural simulation architecture organized around a unified, independently checked representation, retaining Brian2's modeling frontend while rebuilding downstream execution. AtlasIR exposes scheduling, effects, state domains, events and identities so CPU, GPU, browser and MPI implementations can apply target-specific strategies within explicit contracts. The measured implementation improves CPU runtime and native memory in a full-connectome cohort, completes a large microcircuit under a laptop memory constraint, executes a 24.1-billion-synapse network across four nodes, and reaches 860 million neurons with 860 billion connections in a five-point synthetic weak-scaling study on up to 30 hosts and 240 physical worker cores. Browser experiments demonstrate portable reference execution and a separate full-connectome AOT application. Supported MPI plasticity and boundary recovery extend stateful execution; a bounded native-training plan family adds explicit forward/reverse execution under declared gradient contracts.

Execution plans connect this representation to inspectable implementation choices, and verified GPU calibration can select faster replays within a bounded policy space. Evaluation also identifies limits: calibration can dominate execution, GPU benefits vary by workload, collectives restrict measured rank scaling, and lower native recording memory can coexist with higher total process memory. These findings locate the contribution in concrete execution and resource mechanisms. The integrated development tree provides a common implementation and workflow base, while the remaining validation concerns joint release qualification, broader comparative efficiency, learning scalability, and scientific applications requiring the capacities enabled by the new engine.

## Code and data availability

The accompanying evidence package contains a source/evidence map, machine-readable figure inputs and deterministic figure-generation scripts. The supplement identifies source snapshots, model and dataset provenance, reproduction commands, and retained report records. Large arrays and some historical artifacts are held in separate archives identified in those records; they are not included in the manuscript package.

The brian2-atlas implementation is frozen at commit `ae649a244e28d7d4b3eb5df4e35215e14c1378c7` (software DOI: 10.5281/zenodo.23269098). The evidence archive, `biorxiv-v1` (DOI: 10.5281/zenodo.23269145), contains workload scripts, source manifests and experimental records. Both GitHub repositories and both Zenodo archive files are publicly accessible. The pinned PD14 reproduction records a successful rebuild and rerun; the validation index and supplement identify each experiment's source version and validation scope.

The connected-capacity evidence retains historical fixed-host observations, and the weak-scaling evidence retains all five accepted layouts through 860 million neurons, their numerical pilots, terminal resource reports, complete-output audits and source identities. Large input and output arrays remain in the identified remote archives; the retained records include their hashes and admission/acceptance scopes. These isolated experiments do not constitute joint qualification of the current development release.

The conditional resource analysis retains source hashes, explicit array/count formulas and activity assumptions; its script generates Supplement S21 and Supplementary Figure S2 without starting simulations or remote jobs.

## Author contributions, acknowledgements, and disclosures

### Author contributions

Xinjun Li conceived and directed the study, developed the execution architecture and software, designed and coordinated the validation and performance studies, curated the research artifacts, interpreted the results, and prepared the manuscript and visualizations. The author is responsible for the scientific content and final manuscript.

## Acknowledgements

The author acknowledges the developers and maintainers of Brian2, the cited simulation frameworks, and the published models and datasets used in this study. Their contributions are attributed in the references, model descriptions and supplementary provenance records.

## Funding

This research received no external funding. Computational and manuscript-preparation costs were borne by the author.

## Competing interests

The author declares no competing interests.

---

## Appendix A. Feature compatibility across execution targets

**Table 7 | Feature compatibility within the declared simulation contracts.** B means a bounded implemented contract described in the manuscript and supplement; X means outside that target's declared contract; NR means no target-wide qualification is established by the evidence retained for this paper. NR does not assert that an implementation is absent. Every B remains subject to model, precision, schedule, topology and resource admission. CPU reference and AOT are grouped only for these rows; their broader feature sets differ.

| Feature | CPU reference / AOT | CUDA / Metal | Generic WASM | MPI CPU |
|---|---|---|---|---|
| Point-neuron equations and explicit updates | B | B | B | B |
| Clock-driven synaptic state and pre/post plasticity | B | B | B | B (explicit/CSR) |
| Delayed pathways and pending-event queues | B | B | B | B |
| Summed variables | B | B | B | B (post only) |
| Linked variables | B | B | B | X |
| Multiple clocks / non-default scheduling | B | NR | B | X |
| Named custom events / EventMonitor | B | NR | B | X |
| Portable pure-expression Function | B | B | B | X |
| SpatialNeuron / GSL-method interface | B (f64) | X | X | X |
| Python NetworkOperation callbacks | B (start/end) | X | X | X |
| State and spike recording | B | B | B | B |
| Continuation / successful-boundary checkpoints | B | B | NR | B (explicit/CSR) |
| Filesystem-backed CSR topology | B (AOT) | NR | X | B |
| Explicit CPU/GPU rank assignment | X | X | X | B (Metal tested) |
| Cross-node simulation | X | X | X | B |
| General topology growth / repartitioned restore | X | X | X | X |

The GPU column denotes the declared f32 simulation profile; it does not imply f64 equivalence. Generic WASM denotes the portable reference consumer, excluding CPU-only SpatialNeuron/GSL adapters and Python callbacks. Browser batch boundaries preserve state but do not establish Brian store/restore or cross-target checkpoint interchange. Recording support does not imply every monitor source, variable, clock or slot is accepted.

MPI uses a shared clock and restricted event slots. Mutable synaptic state, segmented execution and successful-boundary recovery apply to admitted explicit/binary-CSR models; the large fixed-total procedural path remains a single tick-zero activation. General structural plasticity, dynamic partition migration and recovery from failed ranks are outside the contract. WebGPU, model-specific WASM AOT, mixed-device MPI and native training retain the separate scopes in Table 1 and Supplements S7/S11/S12/S16. The matrix describes target-specific feature contracts and their separately identified validation scopes; it does not establish full Brian2 API coverage or joint qualification of every feature combination.

## S13. Additional dendritic CPU evidence

The workload source is the contextual-dendritic-gating repository snapshot dbb77525f2662199544f5a0d3dcc9c18b0e1c853. The inspected model guide is retained as dendritic_review.md. The archived nonlinear and linear single-neuron aggregates retain their original implementation identities. Their exact hashes, numerical-gate fields, timing definitions and median-ratio checks appear in evidence.json.

Both single-neuron aggregates declare the same host, process affinity, protocol and topology within each campaign, Cython/Rust/Rust/Cython order, warmup exclusion and unprofiled samples, with ten measurements per backend. Per-activation comparisons of feedforward/silent weights pass at rtol 1e-12 and atol 1e-14. Their Cython/Rust ratios are 12.491379440383696 and 14.752074507741593. These selected-state checks do not establish all-trajectory equality. Construction, export, compilation, checkpoints, result dumps and plotting remain outside the specified primary intervals. Other upstream figure families retain unresolved scientific gates and are not claimed as complete paper reproduction.

The retained adapted model consists of sixteen source files with tree digest 89cb7eba9d314176f61e05795c1f6aebf08cb84825df46a59b0fb60d0ae2b108; the run metadata records model-source revision 73feb595ede908a368947d932055dc0a4e1b3817. The upstream reference snapshot cited above and this retained adapter are distinct provenance records. The common runtime snapshot does not replace the external model-source identity.

The frozen-source network retest is bound to common-source identity cc67a82bfeed8ce850c264b4ace7924b9b56be116b4132bb3e4eed9e2b020112 (manifest, freeze); its regenerated model SHA-256 is 52a5202630b2689abda2eabfa87fd43b32ca15041266af0f2613d4b302378490. The frozen topology is unchanged. The matching independent Rust validator is built from source matching this snapshot; Cargo.lock, rustc 1.98.1 and the existing Zig 0.16.0 C/C++ environment are retained. Activation order is Cython, cache off, cache on, cache on, cache off, Cython, with one discarded warmup and three unprofiled measurements each. The production aggregate and same-source ablation both pass protocol, affinity, topology and all 21 state-array gates. Their common Cython median is 9.796825735 s; Rust medians are 11.564384068 s with the cache disabled and 7.631756929 s with production defaults. All four newly exported model hashes match, and all complete native state/event dump hashes match. The production Cython/Rust ratio is 1.283692055; cache reuse reduces runtime by 34.006%.

**Table S5 | Same-source endpoint-cache ablation.** Both policies use the same frozen source, regenerated AtlasIR, full topology, numerical precision, compiler flags and Cython control. Each row reports six measured observations per backend over 100 ms model time; all reported times are steady-state simulation/required-recording medians in seconds. Ratios are Cython divided by Rust. Cache-off execution is an explicit experimental ablation; endpoint reuse is enabled by production defaults.

| Rust cache policy | Samples per backend | Brian2 Cython (s) | Rust AOT (s) | Cython / Rust |
|---|---|---|---|---|
| Disabled (ablation) | 6 | 9.796826 | 11.564384 | 0.847 |
| Enabled (production default) | 6 | 9.796826 | 7.631757 | 1.28 |

The cache is selected only for pure float64 endpoint expressions independent of the accumulated destination and edge-dependent or rebound inputs. It is refreshed in each original activation, after scalar statements, without reordering reductions. The three NMDA current sums have source-level evaluation counts of 1,112,596 edges × 1,000 ticks versus three endpoint vectors of 2,400 values × 1,000 ticks. The active regression checks pre/post sums, offset subgroups, multiple clocks, plastic event updates and a continued-run scalar/state change; 105 source and 42 target spikes are nonzero, NumPy comparisons pass, and cached/uncached outputs are bit-exact. The retained test log records 29 passed related tests. The complete evidence audit binds reports, raw-array hashes and source identities.

Table 5 reports default implementations. The network short protocol uses the source manifest linked above. The two single-neuron cohorts use the source versions identified in their respective evidence records.

## S21. Conditional resource analysis at an 86-billion-neuron reference count

### S21.1 Scope and workload mapping

This analysis concerns the synthetic recurrent E/I workload in S20, with reference-f64, expected mean indegree K = 1,000, dt = 0.1 ms and 100 ms model time. N = 86 billion therefore gives E = KN = 86 trillion connections. It does not represent reconstructed human anatomy, a validated whole-brain dynamical model, or a new execution. The machine-readable analysis, scenario table and analysis script bind arithmetic and assumptions to the accepted observations and frozen source archive. The twelve inspected source files are retained and hash-verified against the experiment catalogue.

Preserving the three-host/86-million-neuron block gives H = 3N/(86 million) hosts and R = 8H ranks/physical worker cores. The endpoint extended by a factor of 100 maps to 3,000 hosts and 24,000 physical worker cores, with 24,000 populations and 576 million population-pair projections. This is a conditional local-work allocation, requiring bounded indices, buffers, reporting and recording. It is not the host requirement of the current implementation, a guarantee that the tested host memory suffices, or a prediction of runtime. Graph realizations and activity may change with scale. The measured 1% launch time is not extrapolated.

### S21.2 Source-derived memory accounting

For one rank owning n neurons and e incoming edges, the retained immutable synapse arrays use 20e bytes: target indices u32 (4), weights f64 (8) and per-edge delay ticks usize (8) on the tested 64-bit target. Local e is approximated by Kn in the conditional table; largest-remainder projection allocation can slightly perturb individual local counts, and no full-scale edge-count matrix is constructed here. Temporary original-edge identities are used during initialization, then dropped for this deterministic additive pathway. Runtime neuron state uses 33n bytes: v/current (16), lastspike (8), refractory-until (8) and a byte flag (1). These exclude allocation capacity, queues and metadata.

There is one dense source-offset array per incoming projection. Summing source populations gives 8(N + P) bytes per rank, where P = R is the population count; the eight ranks on a host therefore retain 64(N + P) bytes. At the full reference count this is 640.750 GiB per rank and 5.006 TiB per host before other arrays. This replication survives adding hosts. Dense index construction also uses a temporary 8n_source-byte cursor for one projection, and temporary 8e_projection-byte original-edge identities coexist with weight/delay construction. They are lifetime-specific temporaries and are not summed across all projections as simultaneous allocations.

Spike-receive capacity requires a separate qualification. `slot_codegen.py` flushes consecutive producers before their summed population count would exceed 2^31 - 1; `exchange_spike_batch` then checks the batch capacity and resizes one reusable Vec. For this workload every individual population remains below the limit. At tested N the single batch has capacity N; above that range, the existing generator splits batches. Thus 8N is not the full-scale receive-buffer formula. The element payload has an upper bound of 8(2^31 - 1) bytes, just below 16 GiB per rank and 128 GiB per eight-rank host, excluding Vec over-allocation and MPI scratch. Every proposed schedule and count/displacement still needs qualification. Batching limits the buffer; it does not remove global spike dissemination or establish efficient communication at 24,000 ranks.

**Table S4 | Full-reference conditional resource components.** Array sizes are logical element payloads on a 64-bit target. They are not a predicted process RSS/cgroup peak, and rows belonging to different lifetimes must not be summed as a simultaneous requirement. Root recording and output estimates assume the observed 1% activity per neuron persists for 100 ms. “Unchanged dense” terms identify costs to redesign, not a feasible execution configuration.

| Component / condition | Full-reference quantity | Interpretation |
|---|---|---|
| Neurons / connections | 86 billion / 86 trillion | Count-normalized synthetic workload |
| Fixed local-work allocation | 3,000 hosts / 24,000 cores | Conditional allocation, not current host requirement |
| Largest local population / incoming edges | 3,621,053 / approximately 3,621,053,000 | Largest-remainder allocation can perturb edge totals |
| Largest host's immutable synapse arrays | 536.206 GiB | 20 bytes per edge; other memory excluded |
| Immutable synapse arrays across ranks | 1.720 decimal PB | 20E bytes; not total cluster memory |
| Largest host's local neuron arrays | 0.885 GiB | 33 bytes per local neuron |
| Unchanged dense source CSR | 640.750 GiB/rank; 5.006 TiB/host | Global source rows replicated per rank |
| Batched spike-receive element bound | <16 GiB/rank; <128 GiB/host | Existing signed-count-aware generator; capacity overhead excluded |
| Root counts / collected final neuron arrays | 640.750 / 2002.344 GiB | Concentrated recording and post-run collection |
| Root logical spike history | 810.215 GiB | 8 bytes per spike; Vec capacity can exceed this |
| Scientific binary payload | 4.578 decimal TB | Final arrays and two spike streams; excludes reports/headers/final-fired/traces |
| Population-pair projections / dense report entries | 576 million / 55,296,000,000,000 | Four scalar entries per projection/rank in each report copy |
| Binary topology recipes in all rank shards | 331.776 decimal TB | 24 bytes/projection/rank, before neuron input/headers |

## S21.3 Activity, queues and recording

For roughly stationary activity f Hz and mean delay d = 1.5 ms, a logical pending-edge estimate is 4efd bytes per rank, because this experiment stores u32 pending edge indices. For the largest host, f = 1, 12.645 and 40 Hz give approximately 0.161, 2.034 and 6.434 GiB of live queue entries. These are activity scenarios, not capacity reservations or hard upper bounds. Transients, heterogeneous delays, ring-slot capacities and allocator growth matter: the accepted 1% run retained up to 1.728 GiB of queue capacity per rank. Averages cannot establish a transient peak.

All ranks receive the spikes and retain population fired/last-fired vectors for canonical event delivery. Their logical global storage depends on activity; a stationary scenario is approximately 16Nf dt bytes per rank for two usize vectors, before retained peak capacities. Communication includes global allgather/allgatherv, validation, sorting and scattering. At endpoint activity S/N = 1.264479571 spikes per neuron per 100 ms, full-reference payload delivery is 8S = 869.962 decimal GB per rank over the run;

aggregate recipient payload R(8S) is 20.879 decimal PB. This is logical recipient payload, not measured fabric traffic or a bandwidth/runtime forecast; MPI algorithms, same-host sharing and network topology alter physical traffic.

Rank zero retains counts (8N bytes) and one compact spike history (8S logical bytes). Post-run collection adds v/current/lastspike/flag arrays (25N bytes) on that rank, with temporary conversion/receive buffers during each population's collection. For unchanged activity, these selected root arrays total 3.372 TiB, excluding its synapses, CSR, queues, capacity overhead and report strings. The observed 1% history allocation was 15.000 GiB versus 8.102 GiB of logical records, illustrating why payload is not peak memory. Distributed/streamed recording and collection are conditions for avoiding this central concentration.

The two scientific binary files store final arrays and both logical spike streams. A selected payload formula is 33N + 16S bytes. At unchanged endpoint activity it gives 4.578 decimal TB for 100 ms. Trace samples, final-fired vectors, headers and provenance reports are additional. Files of this size exceed the tested 64 GiB per-file guard; output sharding/streaming or newly qualified finite limits would be required. This is not a forecast of total output, and it is not extrapolated to longer model times or different activity.

### S21.4 Projection metadata, construction and output provenance

The unchanged population-pair representation has $P^2$ projection recipes/objects on every rank. Binary shards serialize edge-count/seed/pending-count words for each recipe, even on non-owner ranks: at least $24P^2$ bytes per shard, giving 13.824 decimal GB per shard and 331.776 decimal TB across all shards, before population inputs. The four per-neuron instance columns alone contribute 4N initial values and 25N binary bytes. They exceed the experiment's four-billion-value admission policy at full reference count. The frozen experiment driver admits only the five tested host counts and its opt-in 860-million-neuron ceiling; these and the 64 GiB IR limit are finite tested policies, not full-reference qualification. The endpoint model JSON was 53,438,046,766 bytes; neither its frontend peak nor compilation/preparation time is fitted or linearly forecast. Lazy/shared recipes, streamed instances and newly qualified finite admission limits would be needed.

Projection reporting already batches 128 projections: the gathered packet is at most 640R u64 values, 0.114 GiB at R = 24,000. However, the reporting function appends the complete dense report string on every process. It retains two integer rank statistics, one build-time value and one CSR-byte value for every projection/rank pair: $4P^2R = 4P^3$ scalar entries per report copy. At full reference count there are 55,296,000,000,000 entries; even a conservative two bytes per entry corresponds to at least 110.592 decimal TB (100.583 TiB) per serialized copy, before other fields. The current formatting is larger. This is a combinatorial lower-bound warning for an unchanged representation, not a proposed memory allocation. Root writes this provenance into both runtime and summary JSON. Sparse owner records, aggregated counters and bounded streaming would need to replace this path. Simply multiplying the observed complete output by 100 would miss this growth.

## S21.5 Measured range and conditional extension

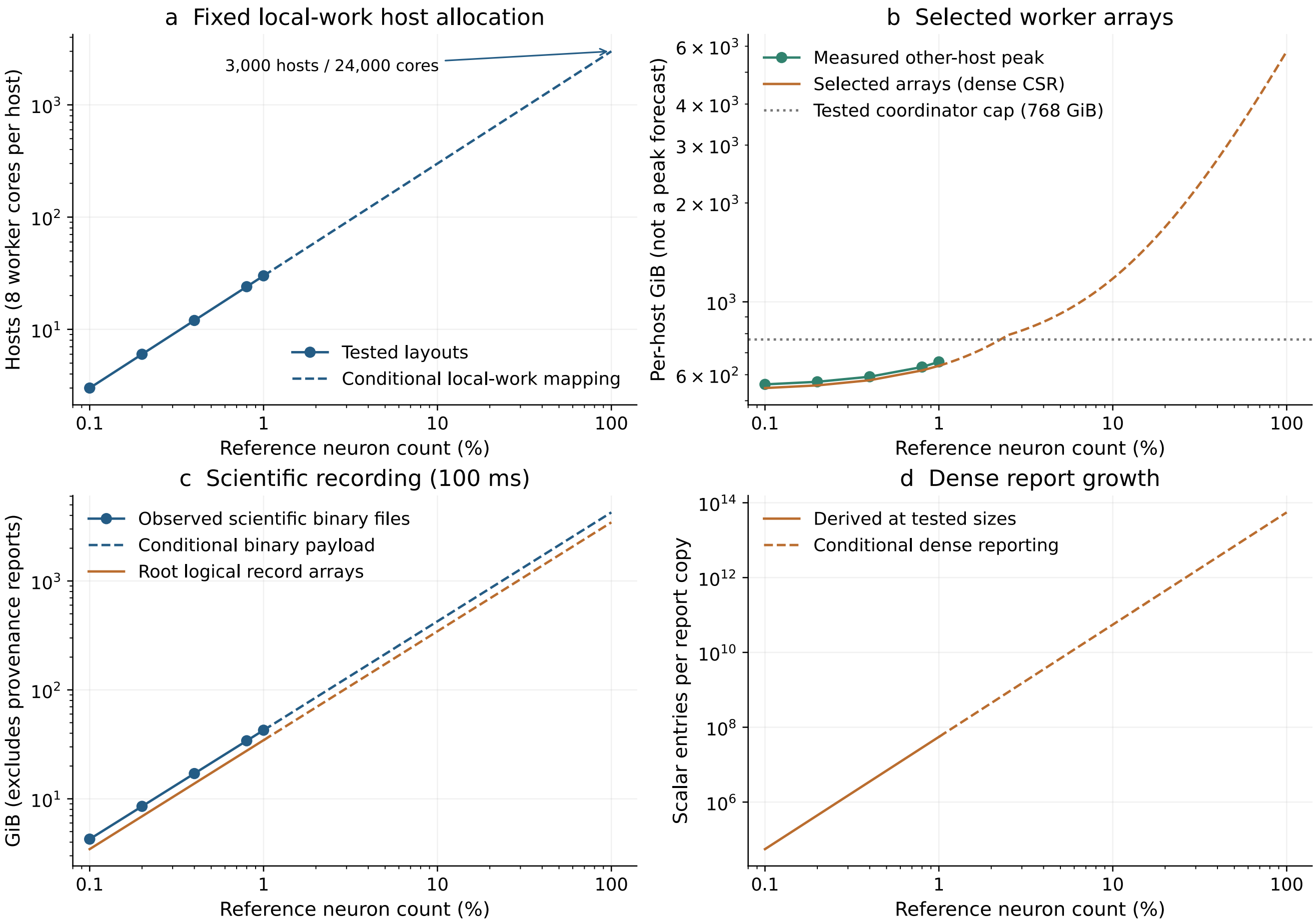

**Supplementary Figure S2 | Source-derived conditional resource analysis.** Solid segments describe tested layouts, observed memory/output, or source-derived quantities at those layouts; dashed segments beyond 1% are conditional estimates/bounds from arithmetic, not executions. (a) Fixed local-work host allocation. (b) Other-host measured peaks and selected arrays retaining dense CSR and the batched receive bound. Selected arrays exclude state, queues, recording, report metadata, allocator and cache; they do not predict peak memory. (c) Scientific binary output and root logical record arrays under fixed endpoint activity and 100 ms, excluding provenance reports. (d) Projection-by-rank report entries under unchanged dense reporting. Axes are logarithmic. No runtime fit, uncertainty band, full-scale host admission or actual full-scale capability is implied.

## S21.6 Conditions for further extension

The tested local-work distribution provides a basis for further distributed capacity studies. Extending it requires reducing global source-index replication, preserving count-safe batched communication while qualifying thousands of ranks, bounding projection metadata/reporting and frontend construction, and distributing recording/result output. New index/layout choices must preserve source/edge identities and event ordering, pass independent numerical gates, and be measured for memory and communication at intermediate scales. The selected array payloads and buffer bounds here are not sufficient evidence that any proposed redesign fits a given host configuration.

This study was self-funded. Execution at the full 86-billion-neuron count was not evaluated within the available computational resources. Resource arithmetic identifies a conditional allocation and specific implementation changes for further investigation; it does not establish full-scale execution feasibility or imply that additional machines alone suffice.

## Supplementary Figure S1 · a | Recorded FlyWire circuit activity

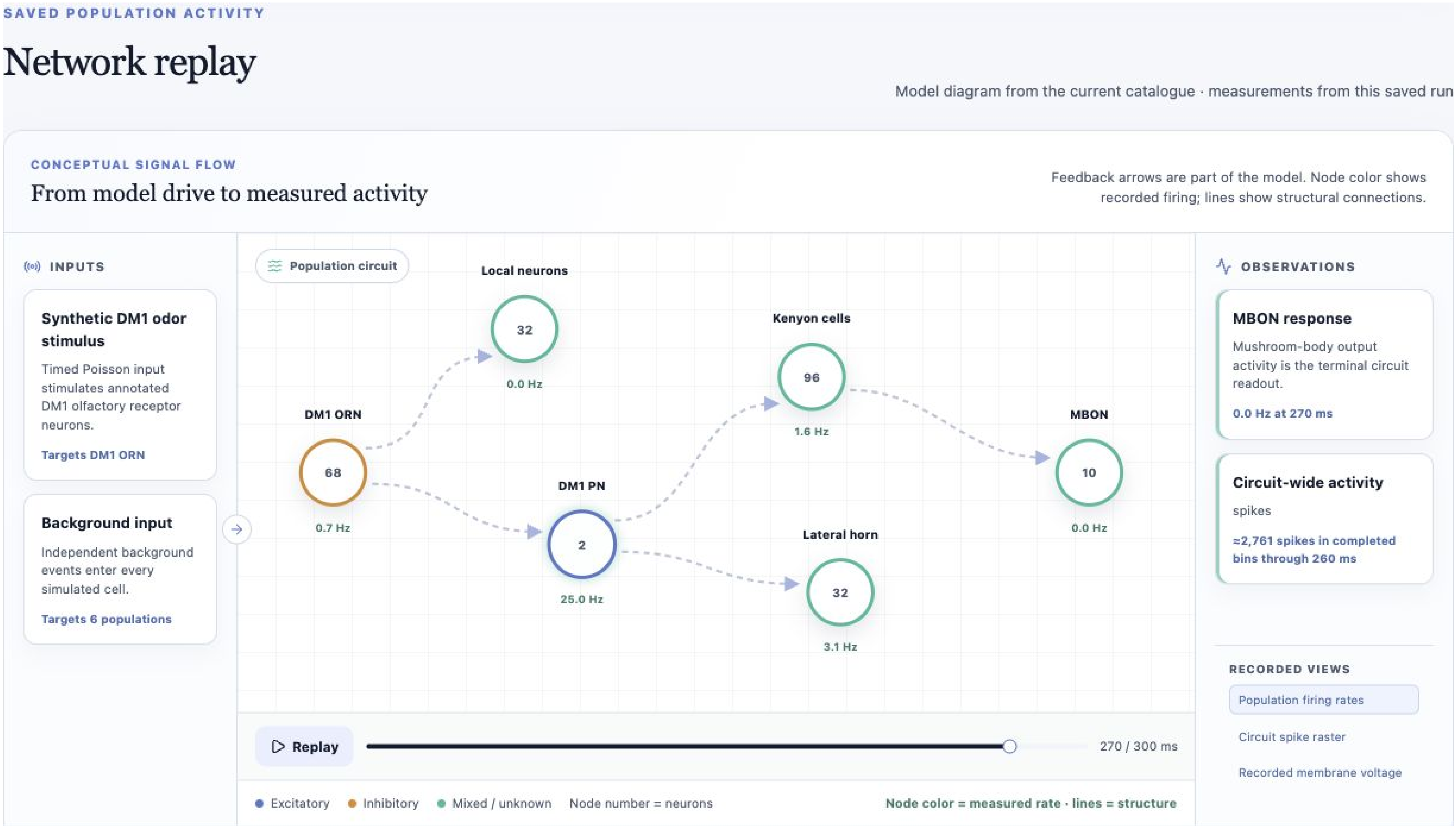

Saved Browser WASM run of FlyWire DM1 (dm1-240, reference-f64): 240 neurons, 300 ms of biological time and 2,796 recorded spikes in the full run. Replay is positioned at 270 ms. The current catalogue supplies the six-group network layout; colors and displayed rates come from saved measurements. This is recorded activity replay, not a new simulation or the AtlasIR schedule.

## Supplementary Figure S1 · b | AtlasIR model draft and logical execution cycle

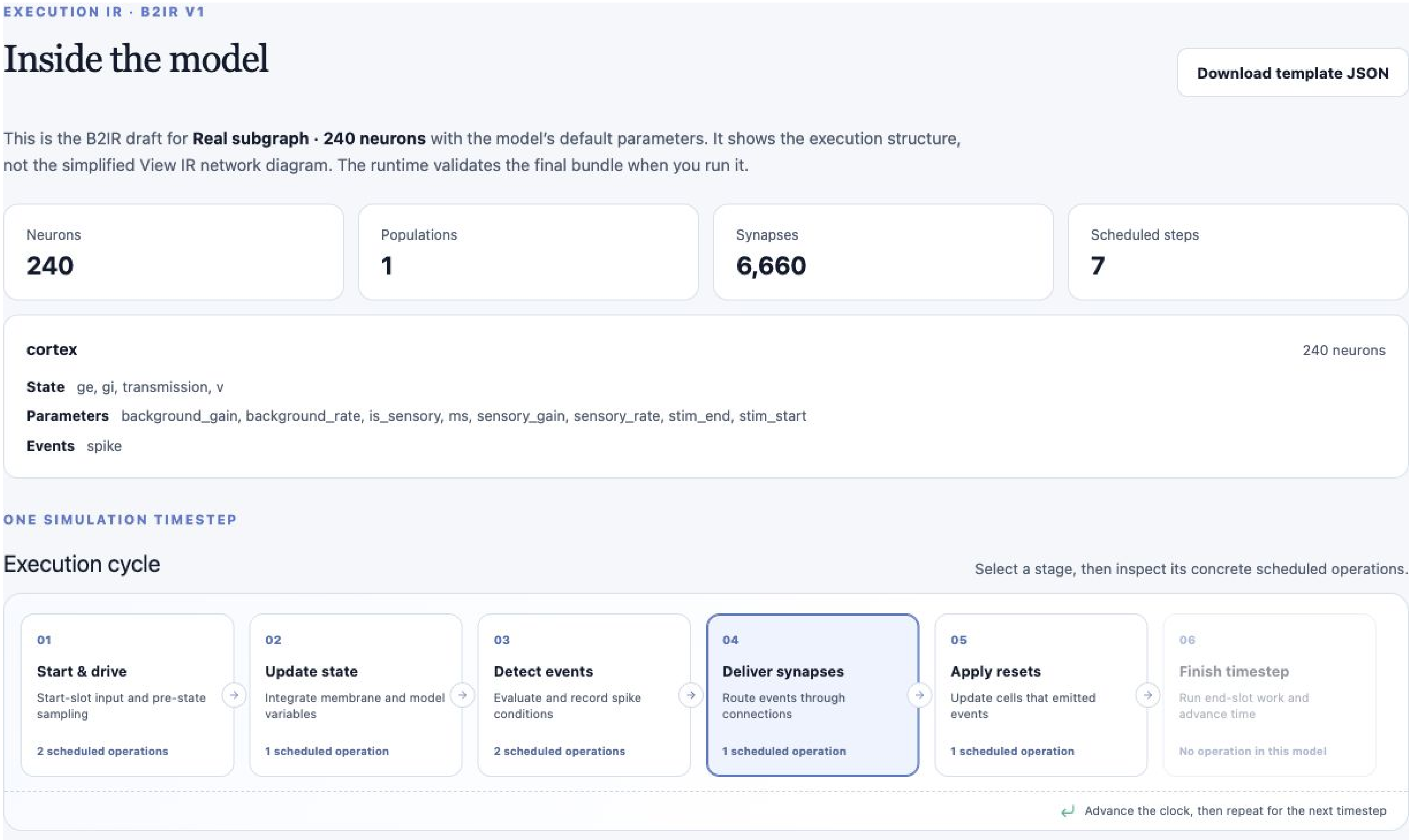

Corresponding online FlyWire DM1 model-page AtlasIR draft, at the 240-neuron scale with default parameters: one execution population, 6,660 synapses and seven scheduled operations. This draft is not the archived execution artifact of panel a. Biological groups in the network diagram and execution populations in AtlasIR serve different purposes. The runtime validates the final bundle on execution.

## Supplementary Figure S1 · c | Synaptic delivery, dependencies and operations

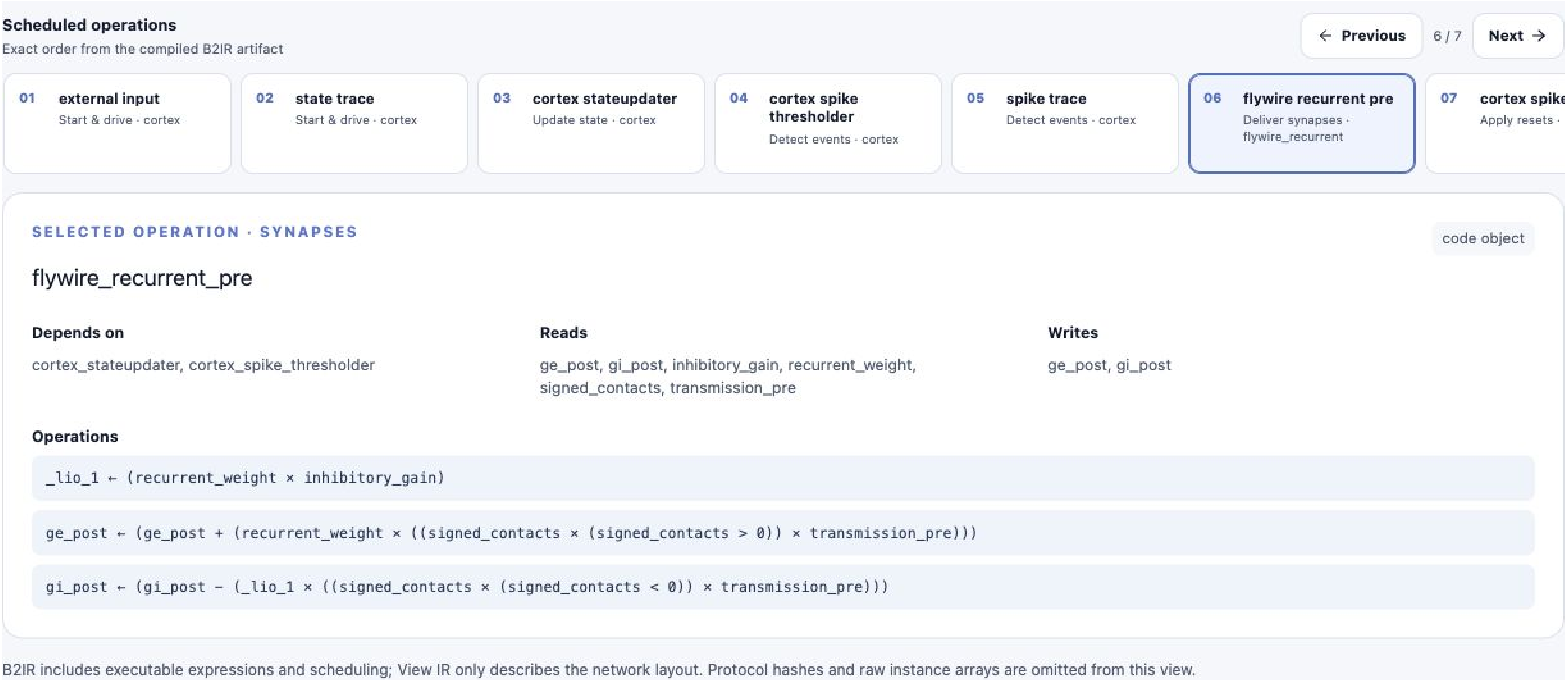

The selected **flywire_recurrent_pre** node depends on the state updater and spike thresholder. It reads postsynaptic conductances, inhibitory gain, recurrent weight, signed contacts and presynaptic transmission, and writes **ge_post, gi_post**. The displayed operations separate excitatory and inhibitory conductance updates. This is logical schedule inspection; the complete target-specific physical plan is not shown.